\documentclass[usenatbib,useAMS]{mn2e}
\usepackage{times}
\usepackage{epsf}

\newcommand{\mtot}{\relax \ifmmode M_{\rm tot}\else $M_{\rm tot}$\fi}
\newcommand{\Reff}{\relax \ifmmode R_{\rm e}\else $R_{\rm e}$\fi}
\newcommand{\SBe}{\relax \ifmmode \langle SB_{\rm e}\rangle \else $\langle SB_{\rm e}\rangle$\fi}
\newcommand{\mB}{\relax \ifmmode M_{\rm B}\else $M_{\rm B}$\fi}
\newcommand{\ReB}{\relax \ifmmode R_{\rm e,B}\else $R_{\rm e,B}$\fi}
\newcommand{\SBeB}{\relax \ifmmode \langle SB_{\rm e,B}\rangle \else $\langle SB_{\rm e,B}\rangle$\fi}
\newcommand{\mD}{\relax \ifmmode M_{\rm D}\else $M_{\rm D}$\fi}
\newcommand{\muo}{\relax \ifmmode \mu_{\rm 0}\else $\mu_{\rm 0}$\fi}
\newcommand{\h}{\relax \ifmmode h\else $h$\fi}
\newcommand{\db}{\relax \ifmmode D/B\else $D/B$\fi}
\newcommand{\bt}{\relax \ifmmode B/T\else $B/T$\fi}
\newcommand{\nb}{\relax \ifmmode n_B\else $n_B$\fi}
\newcommand{\inc}{\relax \ifmmode i\else $h$\fi}
\newcommand{\ellip}{\relax \ifmmode \epsilon\else $\epsilon$\fi}

\newcommand{\magarc}{mag arcsec$^{-2}$}

\newcommand{\vmax}{\relax \ifmmode V_{\rm max}\else $V_{\rm max}$\fi}
\newcommand{\Rq}{$R^\frac{1}{4}$}
\newcommand{\Rn}{$R^\frac{1}{n}$}
\newcommand{\hi}{\mbox{H\,{\small I}}}

\title{Structural properties of discs and bulges of early-type galaxies}

\author[De Jong et al.]{%
Roelof S.\ de Jong$^1$\thanks{E-mail: dejong@stsci.edu}, Luc
Simard$^2$, Roger L.\ Davies$^3$, 
R.~P.\ Saglia$^4$, David Burstein$^5$,\newauthor 
Matthew Colless$^6$, Robert McMahan$^7$ and Gary Wegner$^8$ \\
$^1$ Space Telescope Science Institute, 3700 San Martin Drive,
Baltimore, MD 21218, USA\\
$^2$ Herzberg Institute of Astrophysics, National Research Council of
Canada, Victoria, BC V9E 2E7, Canada \\
$^3$ University of Oxford, Astrophysics, Keble Road, Oxford OX1 3RH\\
$^4$ Max-Planck Institut f\"ur extraterrestrische Physik, Giessenbachstrasse D-85748 Garching\\
$^5$ Department of Physics and Astronomy, Arizona State University,
Tempe, AZ 85287-1504, USA\\
$^6$ Anglo-Australian Observatory, P.O.\ Box 296, Epping, NSW 1710, Australia\\
$^7$ Department of Physics and Astronomy, University of North Carolina, 
CB\#3255 Phillips Hall, Chapel Hill, NC 27599-3255, USA\\
$^8$ Department of Physics and Astronomy, Dartmouth College, Wilder Lab.,
Hanover, NH 03755, USA
}

\begin{document}
\date{\fbox{{\sc Not for Distribution --- Draft:} \today}}

\label{firstpage}

\maketitle

\begin{abstract}

We have used the EFAR sample of galaxies to investigate the light
distributions of early-type galaxies. We decompose the 2D light
distribution of the galaxies in a flattened spheroidal component with
a \citet{Ser68} radial light profile and an inclined disc component
with an exponential light profile. We show that if we assume that all
galaxies can have a spheroid {\it and} disc component, that then the
brightest, bulge dominated elliptical galaxies have a fairly broad
distribution in the S\'ersic profile shape parameter \nb, with a
median of about 3.7 and a sigma of $\sim$0.9. Other galaxies have
smaller \nb\ values. This means that spheroids are in general less
concentrated than the de Vaucouleurs \Rq-law profile, which has \nb=4.

While the result of our light decomposition is robust, we cannot prove
without kinematic information that these components are spheroids and
discs, in the usual sense of pressure- and rotation-supported stellar
systems. However, we show that the distribution of disc inclination
angles is consistent with random orientation if we take our selection
effects into account. If we assume that the detected spheroids and
discs are indeed separate components, we can draw the following
conclusions: 1) the spheroid and disc scale sizes are correlated; 2)
bulge-to-total luminosity ratios, bulge effective radii, and bulge
\nb\ values are all positively correlated; 3) the bivariate space
density distribution of elliptical galaxies in the (luminosity, scale
size)-plane is well described by a Schechter luminosity function in
the luminosity dimension and a log-normal scale-size distribution at a
given luminosity; 4) at the brightest luminosities, the scale size
distribution of elliptical galaxies is similar to those of bright
spiral galaxies, but extending to brighter magnitudes; at fainter
luminosities the scale size distribution of elliptical galaxies peaks
at distinctly smaller sizes than the size distribution of spiral
galaxies; and 5) bulge components of early-type galaxies are typically
a factor 1.5 to 2.5 smaller than the disks of spiral galaxies with a
slight luminosity dependence, while disc components of early-type
galaxies are typically twice as large as the discs of spiral galaxies
at all luminosities.


\end{abstract}


\section{Introduction}

Studying the light distribution of galaxies in order to determine and
compare their structural properties has a long tradition (e.g.,
\citealt{deV48,Fis64,Fre70}). Traditionally, the light
distributions of elliptical galaxies have been fitted by the \Rq-law
profile of \citet{deV48}, while spiral galaxies have been fitted with
exponential light profiles for their discs and often, due to their
similarities to early-type galaxies, \Rq-law bulges (e.g.,
\citealt{deV59,Kor77}). In the recent decade we have seen some shift
from this paradigm. A generalization of the \Rq\ and exponential light
profile laws described first by \citep{Ser68}, resulting in an \Rn\ law,
has more often been fitted to early-type galaxies to account for
observed deviations from the \Rq profiles (e.g., \citealt{Cao93,
  GraCol97}). Based on this development and based on theoretical
predictions for some forms of bulge formation (e.g.,
\citealt{Com90,PfeNor90}), more general forms of light profiles have
been fitted to the bulge light of spiral galaxies as well
\citep{And95,deJ96II,Mac03}. With the realization that many early-type
galaxies have disc-like components
(e.g., \citealt{Ben89,RixWhi90,JorFra94}), bulge/disc decompositions of
elliptical galaxies have been performed, but in general with
\Rq\ bulges and exponential discs (e.g., \citealt{EfarIV}) except for
a few notable exceptions on small samples of galaxies \citep{DOn01,
  Bal04, Gut04}. Here we perform 2D bulge/disc decompositions on the
large EFAR sample of early-type galaxies (\citealt{EfarI}, hereafter
Paper I) using an \Rn\ spheroid and exponential disc light profiles,
determine relations between their structural parameters and their
frequency of occurrence, and compare those to the properties of disc
dominated galaxies.


%
%
%

In an earlier paper in this series (\citealt{EfarIII}, Paper III
hereafter) we presented the luminosity profiles for the 776 galaxies
observed in the EFAR project to which we fitted seeing convolved
\Rq-law bulge plus exponential disc models.  In Paper III, 31\% of the
sample proved to be spiral or barred spiral galaxies and 69\% of the
sample were classified as early-type galaxies.  Of those early-types
18\% were well fit by an $R^{1/4}$ alone, the great majority (70\%)
were very much better fit by the combination of an exponential disc
plus \Rq\ bulge than by a bulge alone.  The remaining 12\% of the
early-type galaxies were classified as cD, without reference to their
profile type, as they were the brightest in their cluster and had
half-total-light radii larger than 10\,kpc.

In Saglia et al. (\citeyear{EfarIV}, hereafter
Paper IV) we showed that the combination of an \Rq-law
spheroid plus an exponential disc is degenerate to an \Rn-law
spheroid, unless one has data for a very large range in surface
brightness and radius.  Given that only 12\% of the EFAR galaxies were
best fitted with an \Rq\ law alone, and that the remainder of the
galaxies were better fitted with an \Rq\ bulge and an exponential
disc, one might wonder about the validity of the \Rq\ law in general.
To address this issue, we have refitted all EFAR galaxies again with
an \Rn-law spheroid and an exponential disc.  If early-type galaxies
are dominated by \Rn\ spheroids, most fits should converge to largely
spheroid dominated systems with small discs.  If the \Rq-law
spheroids plus disc picture is correct, most fits should converge
towards spheroids with S\'ersic parameter $n$ close to 4.  Obviously,
the truth could be somewhere in between with spheroids with S\'ersic
parameter $n$ covering a large range and with a large range in
bulge-to-disc ratios.

We have to realize that even when we find that most early-type systems
have light components that are well described by exponential light
profiles, this in itself will never prove the existence of exponential
discs. First of all, there is no reason why the exponential discs
observed in most spiral galaxies should be present in a similar way in
bulge dominated systems, especially realizing that exponential discs
in spiral galaxies are still poorly understood. Secondly, even if
light is present in exponential distribution, it might not be in a
disc configuration flattened by rotation as discs of spiral galaxies
are normally understood. One can probably make many stable
configurations in bulge-like potentials that have small exponential
like stellar distributions on top of them, without the need of any net
rotation. The best confirmation of exponential light distributions in early
type galaxies being rotationally flattened discs as in spiral galaxies
has to come from measurements of their kinematics.

Even when kinematics is available, it still might be difficult to
reach an unambiguous result. \citet{ScoBen95} made dynamical models
for a small sample of early-type galaxies based on their photometric
bulge/disc decompositions and showed that the observed asymmetries in
the long-slit stellar absorption line profiles were consistent with the
dynamical model predictions.  However, with their data and modelling
technique they could not prove this was an unique kinematic
interpretation.
Better kinematic
constraints on discy components can be obtained with full 2D velocity
profile measurements, such as obtained with the SAURON project
\citep{Fal04,Ems04}. However, so far
these studies have been limited to the high surface brightness central
regions of galaxies.

In recent years a number of studies were performed where the galaxy
samples were large enough and the selection bias understood well
enough that bivariate space density distributions of structural
parameters could be calculated. \citet{Cro01} presented the bivariate
(luminosity, surface brightness distribution) for the 2dFGRS
\citep{Col01}. \citet{Bla03} showed many different bivariate space
density distributions derived from the Sloan Digital Sky Survey (SDSS;
\citealt{Yor00}), and presented some distributions in bins of colour
and concentration index to split their sample in different galaxy
types. \citet{Kau03} derived mass-to-light ratios for the SDSS
galaxies and presented stellar mass bivariate
distributions. \citet{She03} also investigated the bivariate
distribution of structural parameters of different subsets of SDSS
galaxies. However, even though calculated, they did not include the
absolute space densities in their analysis.  While based on large
sample to provide good statistics, in none of these studies bulge/disc
decompositions were performed to derive space density distributions
for the individual components. Such an analysis was performed on the
discs of a sample of about 1000 Sb-Sdm galaxies by
\citet{deJLac00}. However, no such analysis on early-type galaxies has
appeared so far.

A fundamental question for galaxy formation is the origin of the
Hubble sequence. Why do some galaxies have discs and others not?  If
in fact almost all galaxies have a disc component but of varying mass,
luminosity, and size, the question of the origin of the Hubble
sequence takes on a different perspective; in semi-analytic models of
galaxy formation early-type galaxies go through phases where they have
substantial discs which are disrupted and re-built during the cycle of
hierarchical merging.  Determining galaxy disc and bulge parameters
and their bivariate distributions will help us to test and constrain
such hierarchical semi-analytic models of galaxy formation.

In this paper we will use a 2D fitting algorithm to measure the bulge
and disc properties in the EFAR sample of early-type galaxies.  We
estimate the frequency, luminosity and size of discs and bulges and
explore the properties of the discs in the context of S0 and spiral
galaxy discs. 

In the next section we describe our method of 2D model fitting and
parameter error estimation and compare our new results with the
results of Paper III.  In section 3 we characterize the properties of
the discs and their relation to the host bulge component.  Our
conclusions and the implications of this work for models of galaxy
evolution are summarized in section 4.  We use a cosmology of
$\Omega_m$=0.3, $\Omega_\Lambda$=0.7, and $H_0$=70 km s$^{-1}$ Mpc$^{-1}$
throughout.

\section{GIM2D: 2D Galaxy Modelling}
 
We decided to use full 2D modelling of the galaxy images, as this
provides extra bulge/disc separation constraints due to the different
flattening of bulges and discs, sometimes combined with the
constraints resulting from the different position angles of the two
components. Performing the modelling in 2D also allows proper weighting
of data points reflecting uncertainties due to photon statistics, flat
fielding uncertainties, sky level uncertainties and model
imperfections (i.e.\ real galaxies are not as smooth as our models).

To extract the best possible bulge and disc parameters and to estimate
their errors we performed the following steps:

Many of the EFAR galaxies have several repeat observations,
so for the 775 objects more than 2500 images are available. We only
used the $R$-band images, because the $B$-band imaging of the EFAR
sample is incomplete. We determined the seeing on all EFAR images by
finding isolated star-like objects in each of them, then measure the
FWHM of each object by fitting a Gaussian profile, and taking the
error weighted average of the eight objects in each image with the
smallest errors after removing the objects with smallest and largest
FWHM.  At this stage we selected for each galaxy the image with the
best seeing. All galaxies with too few good foreground stars ($<3$) to
make a reliable seeing measurement were removed from the sample (23
galaxies).

SExtractor \citep{BerArn96} was used to create a mask of all objects on the
images, splitting the images of overlapping objects in separate components.
This mask is used by the fitting routine to exclude all pixels with
objects detected except for the pixels with the galaxy of interest and
the sky pixels. A very low level SExtractor mask was also used to
determine the sky level of the image, using a kappa-sigma clipped median
of the pixels deemed free of objects, and this sky level was kept fixed
during the fitting of the galaxy.

The 2D modelling of the galaxy images was performed by the publicly
available GIM2D fitting package (Version 2.2.1), described in detail
in \citet{Sim02}. The spheroidal component of each galaxy was
fitted with a S\'ersic law:
 \begin{equation}
\Sigma_{\rm B}(r) = \Sigma_{\rm e} {\rm e}^{-c([r/\ReB]^{1/\nb}-1)},
\label{expbul}
 \end{equation}
 where the effective radius (\ReB) encloses half the total luminosity
and $\Sigma_{\rm e}$ is the surface brightness at this radius. The 
parameter $c$ is set to 1.9992\nb-0.3271 so that \ReB\ remains the half 
total light radius of the bulge. Better approximations of $c$ were
published after we did our calculations (e.g., \citealt{Mac03}),
but these improvements are only relevant for \nb\ values much less
than 1, which are rare in our sample. The bulge
component was allowed a flattening with ellipticity $\ellip = 1-b/a$, with
$a$, $b$ semi-major and minor axis. The \ReB\ provided by GIM2D is the bulge
half-total-light radius of the major axis, which we multiplied by
$\sqrt{1-\ellip}$ to be consistent with our previous work, which uses
the effective radius on a circular aperture.

The disc was modelled with an
exponential disc:
 \begin{equation} 
\Sigma_{\rm D}(r) = \Sigma_0 {\rm e}^{-r/\h},
 \end{equation} 
which was assumed to be infinitely thin, transparent and have an
inclination $\inc$. The two galaxy components had a common center
which was fitted as well, and an independent flux, here expressed in a
bulge-to-total light ratio (\bt).  Each component has also an
independent position angle. The summed galaxy components were
convolved with a Gaussian with a FWHM equal to the determined seeing,
using an FFT algorithm.  GIM2D fits this large 11-dimensional
parameter space of the two components with the Metropolis algorithm
\citep{Met53}. The Metropolis algorithm is very CPU intensive, but it
tends to be more robust than methods based on gradient searches.  The
Metropolis algorithm is ideal for cases as the current where many fits
have to be performed unsupervised.  The CPU intensiveness was solved
by using 16 fast Linux Dual Processor PCs in a cluster.
 
The fitted scale size parameters in pixels and intensities in counts
were converted to arc-seconds and magnitudes using calibrations
described in Paper III.  To set the nomenclature, the galaxies are
parametrized in the remainder of the paper as follows: for the total
galaxy we have luminosity \mtot, effective (half-total-light) radius
\Reff\ and effective surface brightness within this radius \SBe; for
the bulge luminosity \mB, effective radius \ReB\ and effective surface
brightness \SBeB; for the disc luminosity \mD, scale length \h\ and
central surface brightness \muo.

After all galaxies had been fitted, galaxies with high $\chi^2$
residuals and/or suspicious model subtracted residual images were
identified.  In many of these cases SExtractor had not detected
objects near the nucleus of the target galaxy or done an insufficient
splitting of nearby objects, resulting in incorrect masks and causing
poor fits. For most of these objects we were able to modify the masks
by hand to give satisfactory result, but a number of objects with
bright foreground stars, many bright companions or the occasional
galaxies with double nuclei were removed from the sample at this stage
(20 galaxies).

We used the model subtracted residual images to identify galaxies with
spiral structure. Even though not necessarily spiral galaxies based on
their \bt\ ratio, these galaxies are marked with special symbols in
the diagrams. We will use the term ``non-spiral galaxies'' in the
remainder of this paper for those galaxies that had no clear spiral
structure in the residual images (454 of the 558 galaxies used in the
final selection). The brightest galaxy of each of the clusters with at
least 3 members as identified in Paper VII of this series
\citep{EfarVII} is also marked with a special symbol in the diagrams.

The GIM2D fit package provides 68\% confidence limits on the fitted
parameters, based on the topology of the parameter space that is build
up during the fit process. These errors take only the uncertainties in the
fitted parameters into account, not uncertainties in fixed parameters
like sky level and seeing. As errors in these parameters are often the
main error in the determined bulge and disc parameters \citep{deJ96II},
we repeated all fits with the maximum error expected in our
determination of sky and seeing values.

We estimated the typical error in our sky estimates by dividing the
image in 4 and measuring the sky in each of the quadrants separately
using the same procedure as for the total image (using object masking
and kappa-sigma clipping). The RMS in the 4 sky averages provides
an estimate in the sky level uncertainty on a galaxy size scale. This
is probably a slight overestimate, as most galaxies are near the
center of the frame, so near the average sky of the total frame, while
the sky RMS of the quadrants will be more affected by sky level
gradients across the frame. Most images used have small sky level
uncertainties: 80\% of the galaxies have quadrants errors less than
0.5\%, only 8\% of the galaxies have quadrant RMS values larger than
1\% of the sky, mainly due to sky gradients. We repeated all GIM2D
fits with 1\% sky added and subtracted, and the resulting fit
parameters should be seen as the maximum errors in the parameters
that can be expected to be caused by sky uncertainty.

For each galaxy we repeated the fit with seeing increased and
decreased by 2\% to determine the errors in the fitted parameters
due to seeing measurement uncertainties. Judging from the RMS in the
seeing measurements of different stars in the different frames, less
than 20\% of the galaxies have uncertainties in their seeing larger
than 2\%. Again, these errors should be seen as the maximum error in
the fitted parameters due to seeing measurement uncertainties.

Where appropriate, we show typical errorbars on our graphs, which are
the median errors for all galaxies in the determined parameters, where
the parameter error for an individual galaxy is the maximum error due
to fit parameter space, sky uncertainty, or seeing uncertainty. We
chose the maximum error here, because it is hard to assign relative
weights to the different error sources. To assign proper weights would
require a full set of Monte Carlo simulations, which is prohibited
given the large computation times involved.

We excluded all galaxies with no measured redshift from our analysis,
mainly spiral galaxies which were excluded from the spectroscopy phase
of the EFAR project as being not interesting for the main aim of the
project. The final sample of galaxies used contains 558 galaxies.

\subsection{Testing the 2D fit routine}
\label{sec:testing-2d-fit}

We tested the GIM2D fitting program by performing fits on 500
artificial galaxy images with characteristics similar to our real
data. The images were 500x500 pixels and the pixel size was
0.65\,arcsec. Skylevel was set at 20.5 mag arcsec$^{-2}$. Galaxy
images were created in a Monte Carlo fashion with total magnitudes
sampled linearly between 12 and 16 mag, with \bt\ ratios sampled
linearly between 0.05 and 1, and having bulges with $18\leq\SBeB\leq
22$ and $1\leq\nb\leq 6.5$, and discs with $20\leq\muo\leq23$.  CCD
readnoise and gain were set comparable to most of our observations and
Poison noise was added appropriate for the detected flux level. The
galaxies were convolved with a Gaussian seeing with FWHM of 1.8
arcsec, which is at the high end of the typical seeing of the used
images (90\% with seeing less than 1.8 arcsec).

\begin{figure*}
\epsfxsize=17.5cm
\epsfbox{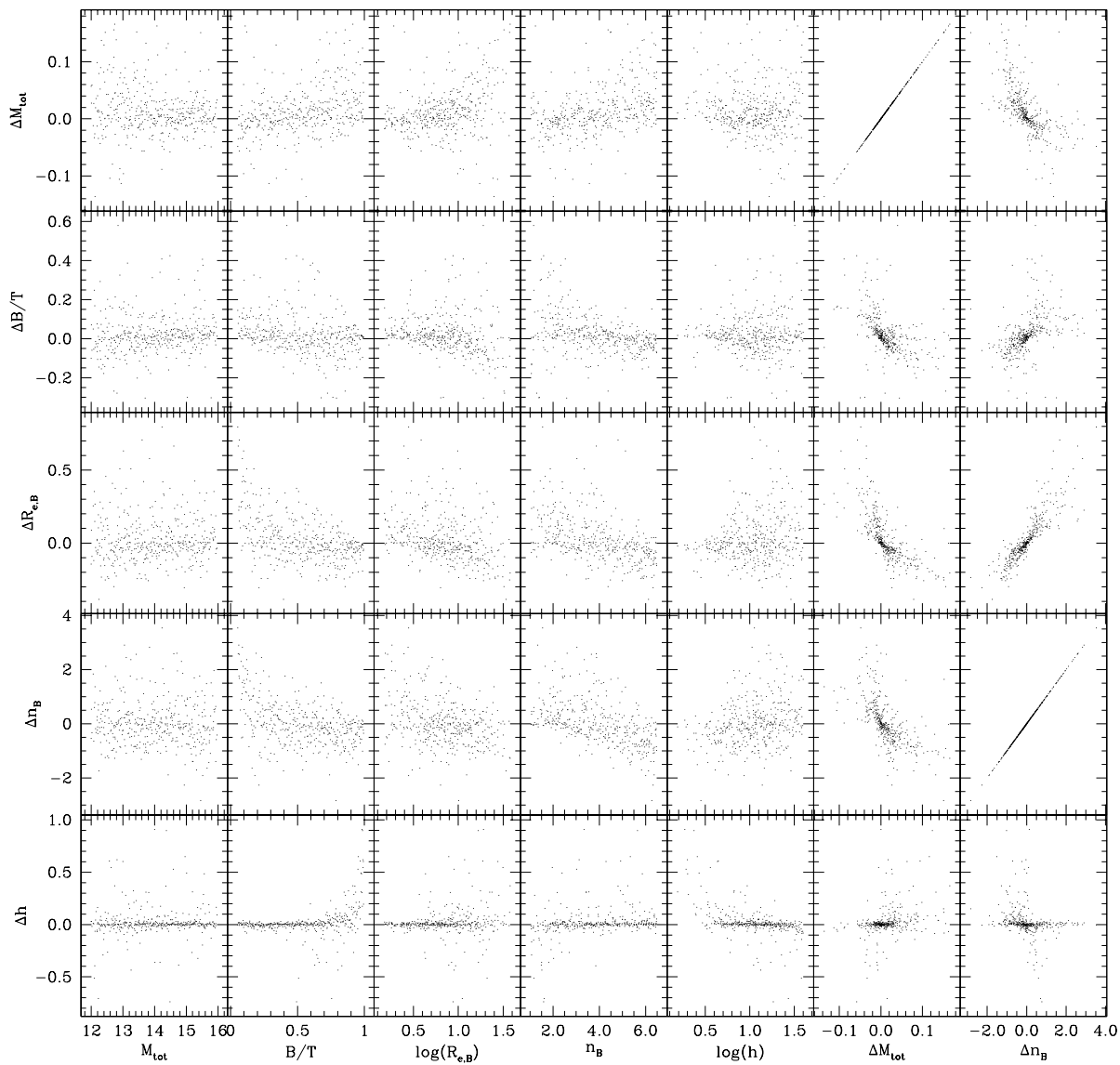}
\caption{Results of GIM2D fits to the Monte Carlo simulated artificial
  galaxy images. The first five columns on the x-axis show the input
  parameters of the galaxy images, the y-axis and the last two columns
  on the x-axis show the differences between fitted and input
  parameters, defined as:
 $\Delta \mtot = \mtot^{\rm fit}-\mtot^{\rm in}$,
 $\Delta \bt = \bt^{\rm fit}-\bt^{\rm in}$,
 $\Delta \nb = \nb^{\rm fit}-\nb^{\rm in}$,
 $\Delta\ReB = \log(\ReB^{\rm fit}/\ReB^{\rm in})$, and
 $\Delta\h = \log(\h^{\rm fit}/\h^{\rm in})$.
}
\label{testg2d}
\end{figure*}

A comparison of the fitted GIM2D parameters with the input parameters
is presented in the first five columns of Fig.\,\ref{testg2d}. All
parameters are reasonably well reproduced, but some systematic trends
can be seen. As could be expected, the largest errors in the bulge
parameters (\ReB, \nb) occur at small \bt\ values, while the largest
disc errors can be found at large \bt\ values and small \h\ values.

Obviously, for real galaxies we do not know the ``input'' parameter
values, but examining the errors in the fitted parameters as function
of the fitted output parameters, shows parameter ranges we should not
trust. We therefore exclude here from the artificial data and later
from the real data all fitted \ReB\ values for galaxies with \bt$<$0.2
or \ReB$<$1.0\,FWHM seeing. Likewise, we exclude all fitted \h\ values
for fits where \bt$>$0.95 or \h$<$1.5\,FWHM seeing. Applying these
cuts results in RMS errors between input and fitted parameters of
order 5\% in \mtot, 10\% in \bt, 30\% in \ReB, 20\% in \nb, and 15\% in
\h. 

It should be noted that some of the errors are strongly correlated, as
shown in the last two columns of Fig.\,\ref{testg2d}. Especially the
bulge parameters \ReB\ and \nb\ are strongly correlated, with some
weaker correlations between these bulge parameters and \mtot\ and \bt.


The \nb\ values show a systematic trend with the input \nb\ in
Fig.\,\ref{testg2d}. This is for a considerable fraction the result of
the limits imposed on the fitted \nb\ values, which were constrained
to $1\leq\nb\leq6.5$. This means that artificial galaxies with \nb\
values close to 1 will have fitted \nb\ values preferentially scattered
to larger \nb\ fit values, and likewise for galaxies with input \nb\
values close to 6.5. A similar effect can be seen in the $\Delta\bt$
versus \bt\ plot. To minimize this effect on real galaxies we
increased the limits to $1\leq\nb\leq8$.

At small \nb\ values a relatively large number of galaxies show
``catastrophic'' errors. These are cases of mistaken identity; the
bulge and disc shapes are so similar that they are swapped. Without
prior assumptions about relative scale sizes or surface brightnesses
this is unavoidable.

\subsection{Comparing profile fitting and 2D fitting}


In Paper III we derived bulge and disc parameters of the EFAR galaxies
using 1D luminosity profiles.  Circularly averaged luminosity profiles
were fitted by the combination of an \Rq\ and an exponential luminosity
light distribution, convolved with seeing, which was also fitted. 
Multiple exposures of the same object were optimally combined and an
optional sky-fitting procedure was used when necessary to correct for
sky subtraction errors. 

We will now compare the Paper III 1D parameters with the current GIM2D
parameters to study the differences between 1D and 2D fitting and \Rq\ 
and \Rn-law bulges. We removed the fit quality $Q$=3 fits from the
Paper III results, as those were considered of poor quality.

\begin{figure}
\epsfxsize=8.2cm
\epsfbox{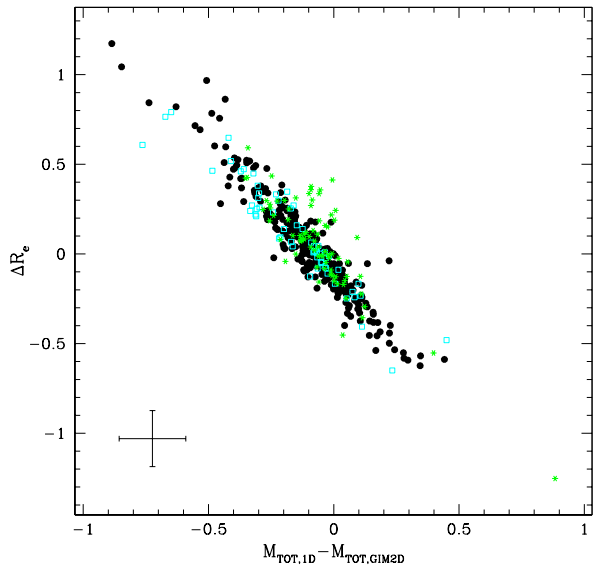}
\caption{The integrated magnitude versus effective radius difference
comparing the Paper III to the current GIM2D determinations. The
effective radius difference is defined as $\Delta\Reff =
2(\Reff$$^{1D}-\Reff$$^{GIM2D})/(\Reff$$^{1D}+\Reff$$^{GIM2D})$. Open
square denote brightest galaxies in EFAR clusters, star symbols denote
galaxies with clear spiral structure in their model subtracted residual
images. The errorbars indicate the median error for all galaxies in
the parameters, where the parameter error for an individual galaxy is
the maximum error due to 2D fit parameter space, sky subtraction
error, and seeing error.
}
\label{cmpMtotRe}
\end{figure}

Figure\,\ref{cmpMtotRe} shows the difference in integrated magnitude
versus the effective radius difference (defined as $\Delta\Reff =
2(\Reff$$^{1D}-\Reff$$^{GIM2D})/(\Reff$$^{1D}+\Reff$$^{GIM2D})$) when
comparing the 1D and the GIM2D results.  The GIM2D magnitudes are
0.06\,mag fainter in the median and agree to
within 0.17\,mag RMS. The effective radii agree to within 28\% RMS, without
taking the strong correlation into account.  The correlated
differences are for a considerable fraction due to the change in bulge
parameter shape allowed by the \Rn\ law.  Even though the flux in the
inner region of the galaxies is well determined, the total flux (and
hence half light effective radius) is determined by extrapolating the
model to infinity.  This is demonstrated more clearly in the next
figure.

\begin{figure}
\epsfxsize=8.2cm
\epsfbox{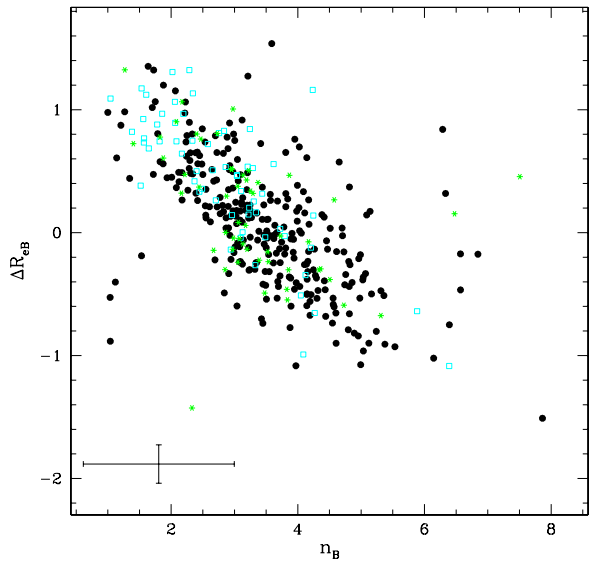}  
\caption{The bulge effective radius difference between 1D Paper III
determinations and  the current GIM2D determinations as function of
bulge shape \nb. The
effective radius difference is defined as $\Delta\ReB =
2(\ReB$$^{1D}-\ReB$$^{GIM2D})/(\ReB$$^{1D}+\ReB$$^{GIM2D})$. Symbols
and errorbars as in Fig.\,\ref{cmpMtotRe}.
}
\label{cmpReBnb}
\end{figure}

Figure\,\ref{cmpReBnb} shows the old versus new bulge effective radius
difference as function of S\'ersic bulge shape parameter \nb. When the
2D fitted \nb\ is getting smaller than the \Rq\ $\nb=4$ value, light
is extending less far and \ReB\ is getting smaller in the \Rn\ law
case than in the \Rq\ law case. The opposite happens when
$\nb>4$. Taking this effect into account, the bulge parameters of the
galaxies agree to within 50\% RMS of each other, a larger
error than for the total galaxy parameters as might be expected.

\begin{figure}
\epsfxsize=8.2cm
\epsfbox{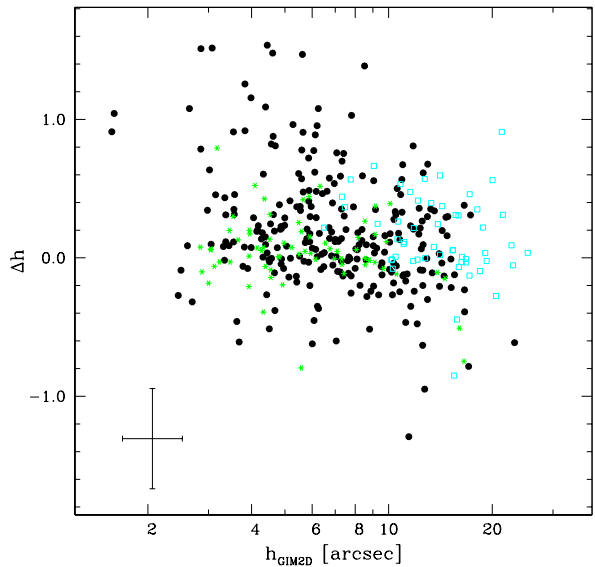}
 \caption{The disc scale length difference between 1D Paper III
determinations and the current GIM2D determinations as function of bulge
shape \nb.  The effective radius difference is defined as $\Delta\h =
2(\h^{1D}_i-\h^{GIM2D})/(\h^{1D}_i+\h^{GIM2D})$, where $\h^{1D}_i =
\h^{1D}/\sqrt{cos(\inc^{GIM2D})}$. Symbols
and errorbars as in Fig.\,\ref{cmpMtotRe}.
}
\label{cmph}
\end{figure}

Figure~\ref{cmph} compares the old and new disc scale lengths. Some of
the extreme outliers are due to the swapping of interpretation of
which component is disc and which is bulge. What used to be a small
disc in a large \Rq-law bulge has become a small \Rn-law bulge with
in general low \nb\ and sometimes vice versa. The disc scale lengths
agree with each other to almost 35\% RMS, which suggests that
many early-type galaxies seem to have a light component that is
robustly fitted by an exponential light profile, to some extent
independent of the shape of the spheroid component. This does not
prove that these are real discs: other radial disc light profiles may
be more appropriate (see e.g.\ \citealt{ScoBen95}) and
kinematics is required to confirm a disc configuration for an
individual galaxy.



In this section we showed that while the 1D \Rq\ and the 2D \Rn\ fit
result in systematically different structural parameters, these
deviations are correlated with the S\'ersic shape parameter. In
determining uncertainties in the parameters we have to take these
systematic chances into account.  Therefore, under the assumption that
the combination of an \Rn\ law spheroid and exponential disc is a
reasonable good description of the intrinsic light distribution of a
galaxy and by giving some more weight to the artificial images test of
Section\,~\ref{sec:testing-2d-fit}, we conclude that bulge and disc
parameters can be determined with a typical 1 sigma accuracy of about
30\%. However, one should keep in mind that some extreme outliers with
much larger errors will be common in large samples. In the remainder
of the paper we will use the GIM2D determined parameters to
investigate correlations between bulge and disc parameters and the
space density of early-type galaxies as function of these
parameters. The fitted GIM2D parameters are available in electronic
format from the Centre de Donn\'es astronomiques de Strasbourg (CDS)
web site.



\section{Properties of Discs and Bulges}

\subsection{Sample selection and selection effects}

The selection of the EFAR galaxy sample has been described in detail
in \citet{EfarI}, and we will repeat here the most important aspects
relevant for the current study. All galaxies were selected with the
intention to be cluster members, with the clusters lying in two
selected regions in the sky which are rich in galaxy clusters. The
target clusters had redshifts in the range from 6000 to 15000
km\,s$^{-1}$. Photographic enlargements of POSS and SERC Sky Survey
plates of the target clusters were inspected for galaxies with
elliptical/S0 galaxy like morphology and a minimum diameter of about
16 arcsec. Morphological selection was done to err on the safe side,
including many spiral galaxies which were identified as such later
with subsequent CCD imaging. The sample is therefore intended to be
complete in large, early-type galaxies (elliptical/S0), but is not
complete in any other galaxy type.  In most of the analysis presented
here, we will only use the galaxies from the EFAR target clusters,
excluding the galaxies from the comparison sample in Coma, Virgo and
the field, as they suffer from completely different selection effects.

To maximize the number of early-type galaxies, galaxies were selected
with rather round isophotes and the sample is therefore rather
face-on. The sample has only eight galaxies with ellipticity
$\ellip$$>$$0.4$ at the outer isophotes.  The distribution of intrinsic
shapes of early-type galaxies is poorly known, and we have therefore
not attempted to correct the bulge parameters to face-on values using
the bulge isophote shapes. However, the disc parameters shown in the
diagrams are face-on values, assuming fully transparent discs.

\begin{figure}
\epsfxsize=8.2cm
\epsfbox{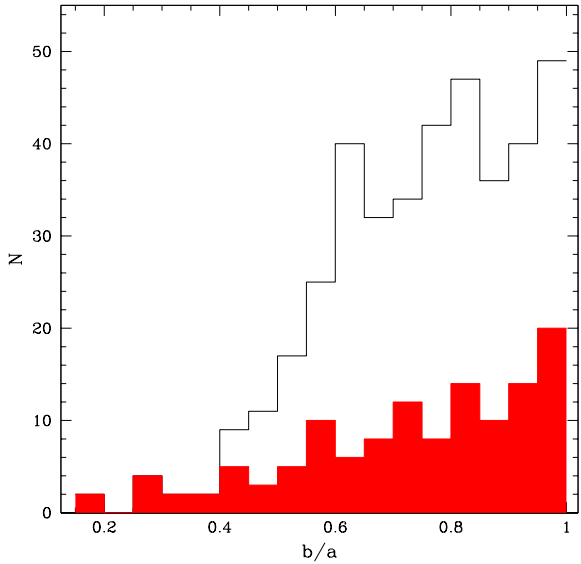}
\caption{The distribution of minor over major axis ratios ($b$/$a$) for
  all non-spiral galaxies (line histogram) and non-spiral galaxies
  with \bt$>$0.7 (solid histogram).
}
\label{bahis}
\end{figure}

While we can not prove that our fitted disc light distributions are
really discs, we can use the minor over major axis ratio ($b$/$a$)
distribution to make a statistical test. For a sample of infinitely
thin discs at random viewing angle the distribution of $b$/$a$ is
expected to flat. We show this distribution for all non-spiral
galaxies in Fig.\,\ref{bahis} (line). The distribution is rather flat
for $b/a>0.6$, but steeply declines for smaller values of
$b$/$a$. This is the result of our selection bias, with very few
galaxies having \ellip$>$0.4 (i.e.\ $b/a$$<$0.6). Assuming that discs
are in general aligned with the spheroid they are sitting in, we
expect the $b$/$a$ distribution to be peaked toward face-on. This
effect will be less for small discs in large bulges and indeed the
distribution becomes relatively more spread out when looking at all
non-spiral galaxies with \bt$>$0.7 in Fig.\,\ref{bahis} (solid
histogram). While not conclusive, this test shows that our fitted
disc components have orientation angles consistent with the random
distribution expected. 

To fully quantify the selection limits of the sample as function of
the structural parameters of the galaxies, we need not only to know
the angular diameter limits, but also the surface brightness at
selection diameter. The average surface brightness at the selection
diameter was 22.05\,$R$-\magarc, with a RMS of 0.5\,$R$-\magarc, as
derived from the 1D profiles of Paper III. This value is used, in
combination with the typical log($D_W$)$>$1.2 diameter limit (Paper I,
see also Eq.\,\ref{D_W}) and the minimum 6000\,km\,s$^{-1}$ distance
limit, to calculate the selection limits indicated in our diagrams
where possible.

\subsection{Galactic extinction and cosmological corrections}

The observed parameters were corrected for Galactic extinction
according to the precepts of \citet{Bur03}, which averages the
estimates of \citet{Sch98} and \citet{BurHei82} with appropriate
offsets.  No attempt was made to correct the parameters for internal
extinction. We used the cluster redshifts as listed in Paper VII to
calculate distances to individual galaxies. Cosmological corrections
were made using $\Omega_m$=0.3, $\Omega_\Lambda$=0.7, and $H_0$=70 km s$^{-1}$ Mpc$^{-1}$,
and K-corrections were made as described in Paper III. All these
corrections are typically small for these galaxies and making
different reasonable assumptions will not affect our results at all.

\subsection{Structural parameter correlations}
\label{sec:struct-param-corr}

We will now investigate correlations in structural parameters of bulge 
and disc. Obviously, many combinations of bulge and disc parameters
can be made, and we will show here only those deemed most
interesting. 

\begin{figure}
\epsfxsize=8.2cm
\epsfbox{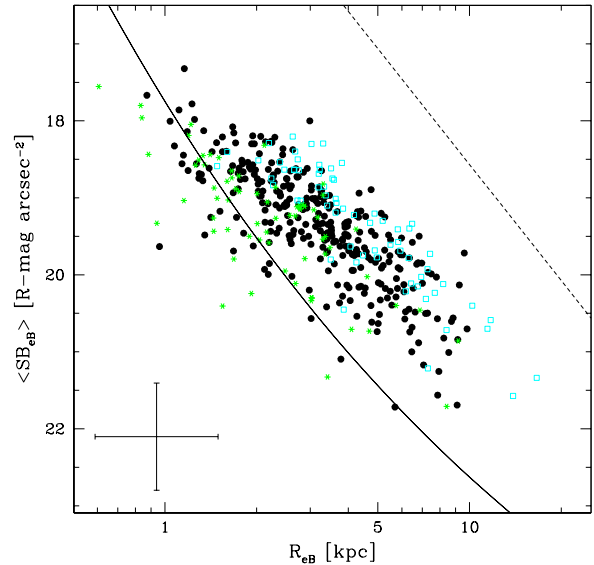}
\caption{Bulge effective radius versus bulge effective surface
brightness. The solid line shows the selection limit for pure \Rq\
profile galaxies at 6000 km\,s$^{-1}$. The dashed line indicates \Rq\
profile galaxies of -25 absolute $R$-mag. Bulges with equal luminosities
will lie on lines parallel to this line. Symbols and erorbars as in
Fig.\,\ref{cmpMtotRe}.}
\label{lgreb_sbemb}
\end{figure}

In Fig.\,\ref{lgreb_sbemb} we show bulge effective radius versus bulge
effective surface brightness for our sample.  There seems to be a
reasonable tight correlation between these two parameters for the
non-spiral galaxies, but we have to be wary of selection effects.  The
solid line in this diagram shows the selection cutoff for galaxies
with an \Rq\ profile dominating at the selection radius, calculated
using the average surface brightness at the selection radius.  The
rather sharp cutoff in the high surface brightness, long scale size
region of the diagram is therefore real; there are no selection
effects acting in this part of the diagram.  This upper limit is
related to the zone of exclusion in $\kappa$-space discussed in detail
by \citet{Bur97}.

At the low surface brightness end of the diagram we have to be more
careful: most of the galaxy distribution seems to lie clearly
separated to the right of the selection limit line, but the selection
line is for a 6000\,km\,s$^{-1}$ cluster, and moving to the right for
the higher redshift clusters.  The ``visibility volume''
\citep{DisPhi83} for these lower surface brightness galaxies is
definitely much smaller than for high surface brightness galaxies,
certainly taking further into account the measurement uncertainty in
the selection diameter (Paper~I).  We will make full ``visibility
volume'' corrections in Section\,\ref{sec:space-dens-distr}, to
calculate the volume density of early-type galaxies as function of
structural parameters.

There are quite a number of galaxies, whose bulges do not follow the
main trend seen for the most non-spiral galaxies in
Fig.\,\ref{lgreb_sbemb}. These bulges lie often to the left of the
selection line, and these galaxies are included in the sample because
their discs make a significant contribution at the selection diameter
or because their profile does not follow an \Rq\ law.  Many of these
bulges, especially in some of the spiral galaxies, may have the more
exponential-like bulges typical of late-type spiral galaxies
\citep{And95,Cou96,Gra01} instead of the more \Rq\-like bulges found
in early-type galaxies.

\begin{figure}
\epsfxsize=8.2cm
\epsfbox{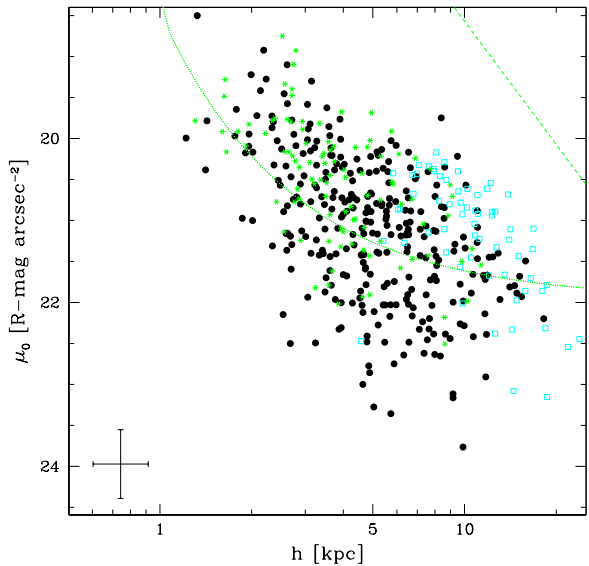}
\caption{Disc scale length versus disc central surface brightness. The 
dotted line shows the selection limit for exponential discs at
$cz=6000$\,km\,s$^{-1}$. The dashed line indicates exponential discs of
-25 absolute $R$-mag. Symbols and errorbars
as in Fig.\,\ref{cmpMtotRe}.}
\label{lgh_mu0}
\end{figure}

\begin{figure}
\epsfxsize=8.2cm
\epsfbox{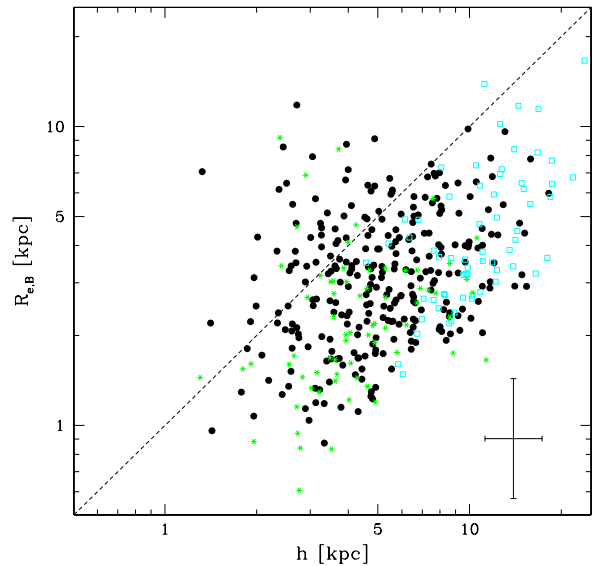}
\caption{Disc scale length versus bulge effective radius. To guide the
  eye, the line of equality is shown as a dashed line. Symbols and
  erorbars as in Fig.\,\ref{cmpMtotRe}.}
\label{lgh_lgreb}
\end{figure}

In Fig.\,\ref{lgh_mu0} we show the disc scale length versus disc
central surface brightness for all fitted galaxies. The dotted line
shows the $cz=6000$\,km\,s$^{-1}$ selection limit for galaxies which
have an exponential light distribution dominating at the selection
radius (i.e. spiral galaxies). Elliptical galaxies will not adhere to
this selection limit, as they were selected on basis of their \Rn\
light distribution. Some interesting trends can be seen in this
diagram, but we have to be wary of more hidden selection effects. At
the very short scale lengths we are close to our scale length cutoff,
as 1\,kpc is of order 2.5 arcsec for $cz=6000$\,km\,s$^{-1}$ and even
less for higher redshifts. The observed limit at the low surface
brightness, small scale length end is more subtle and probably arises
from the minimum contrast needed to see a low surface brightness disc
on top of a bright bulge distribution. Using the relation between \h\
and \ReB\ shown in Fig.\,\ref{lgh_lgreb} discussed in the next
paragraph and the relation between \ReB\ and \SBeB\ shown in
Fig.\,\ref{lgreb_sbemb}, we find that we probe discs with central
surface brightnesses at most about 2 magnitudes fainter than the
\SBeB\ of the bulge. It is not unreasonable to suspect that the low
surface brightness cutoff is caused by lack of disc/bulge
contrast. The only cutoff observed in this diagram that is definitely
real is at large scale length, high surface brightness, as no
selection effects operate in this part of the diagram. This cutoff is
similar to the one observed for spiral galaxies \citep{deJ96III}.

In Fig.\,\ref{lgh_lgreb} we show the correlation between bulge and
disc scale sizes. No obvious selection effects operate in this
diagram, except for a spiral galaxy having a large disc scale length
and a small bulge would probably not have entered the sample. This is
not the case for elliptical galaxies and therefore we have to make
sure that other effects are not causing this correlation, before we
can accept it as real. Our Monte-Carlo simulations showed that our
GIM2D fitting routine is capable to recover models of the full
parameter space presented here, so it is unlikely that it is the
result of a fitting routine artifact. Indeed, we do see a small number
of galaxies with $\ReB/\h>1$ in Fig.\,\ref{lgh_lgreb}, and so the
routine is capable of covering this parameter space in real
data. Still, given the limits on the parameter space for a given
galaxy (not larger than image size, not smaller than seeing), it is
not completely surprising that the bulge and disc scale parameters are
similar in size.

The $\ReB/\h>1$ galaxies can be divided in three groups. A number of
these galaxies are clearly disturbed by dust-lanes or are poorly
fitted due to nearby bright galaxies or stars. We may have a number of
cases of mistaken identity, where bulge and disc are swapped. This is
especially the case when \nb\ is low and the bulge has an almost
exponential light profile. Swapping the bulge and disc parameters on
these galaxies would bring them in line with the general
trend. Finally there are a very few galaxies which seem to have a
genuine small inner light component for their given bulge size, but
these should not be mistaken for the nuclear discs found in HST images
of early-type galaxies, as these are a factor 5--20 smaller yet again
\citep{ScoVan98}. In a number of cases they might be identified
with the inner break radius observed in HST images of early-type
galaxies \citep{Fab97, Tru04}, as shown by \citet{DOn01}. 

\begin{figure*}
\epsfxsize=16.4cm
\epsfbox[50 175 573 408]{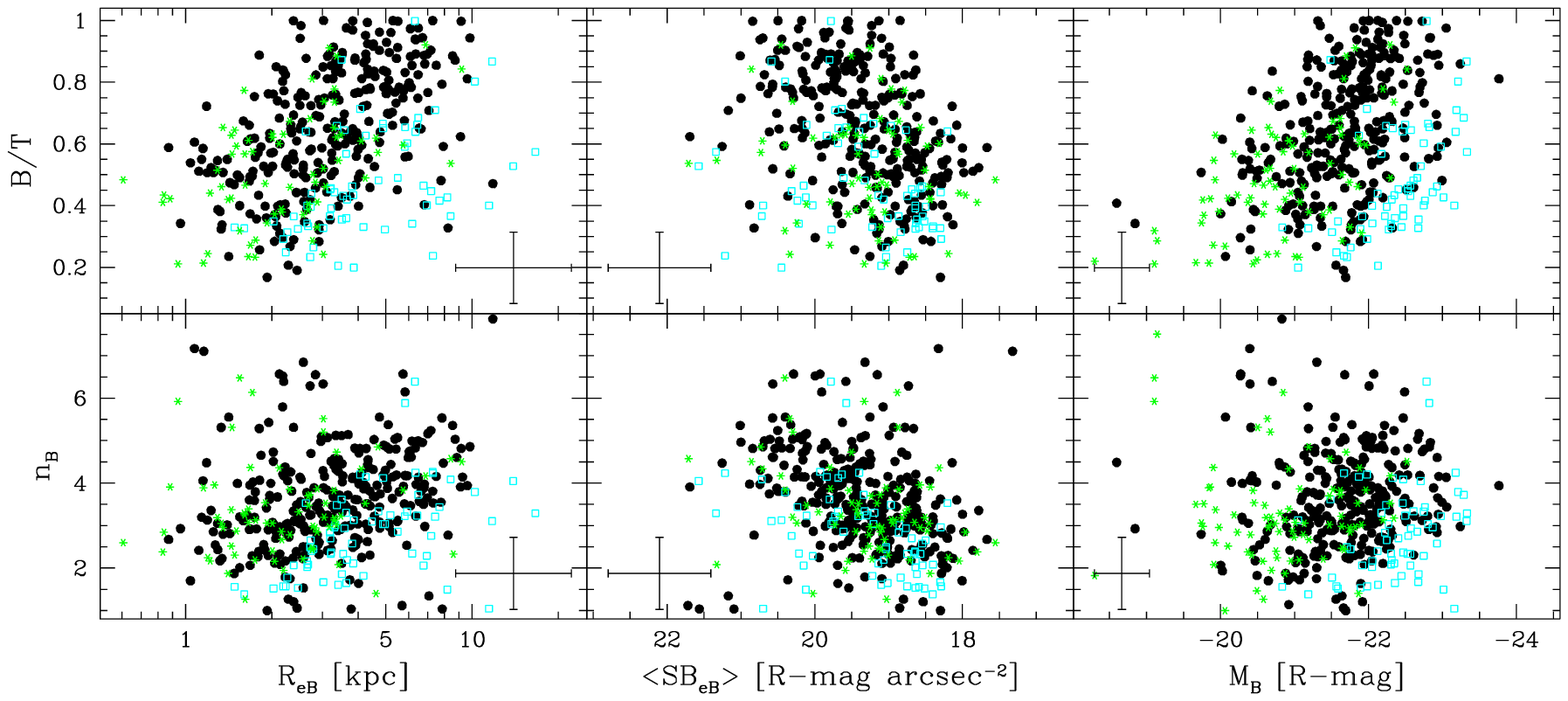}
\epsfxsize=16.4cm
\epsfbox[50 175 573 408]{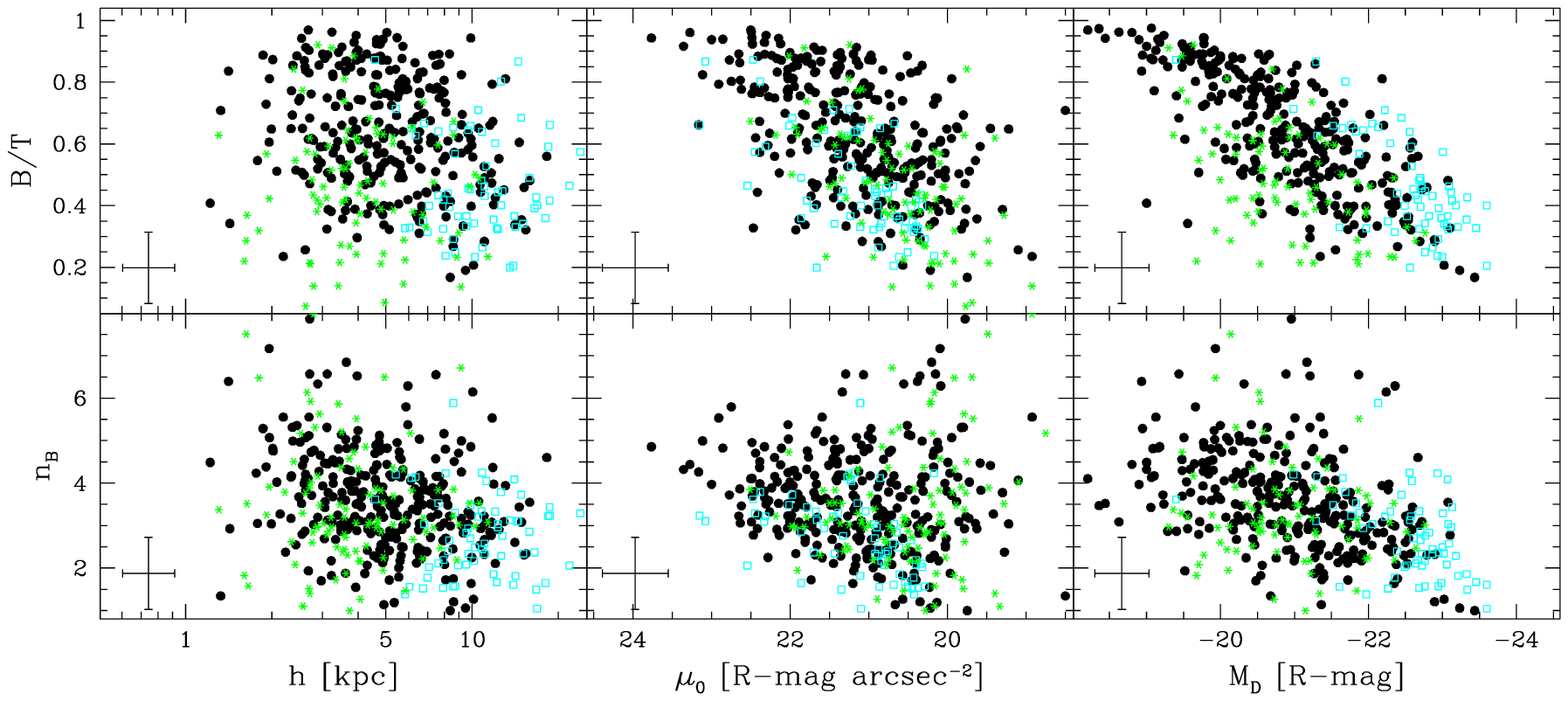}
\epsfxsize=16.4cm
\epsfbox[50 175 573 408]{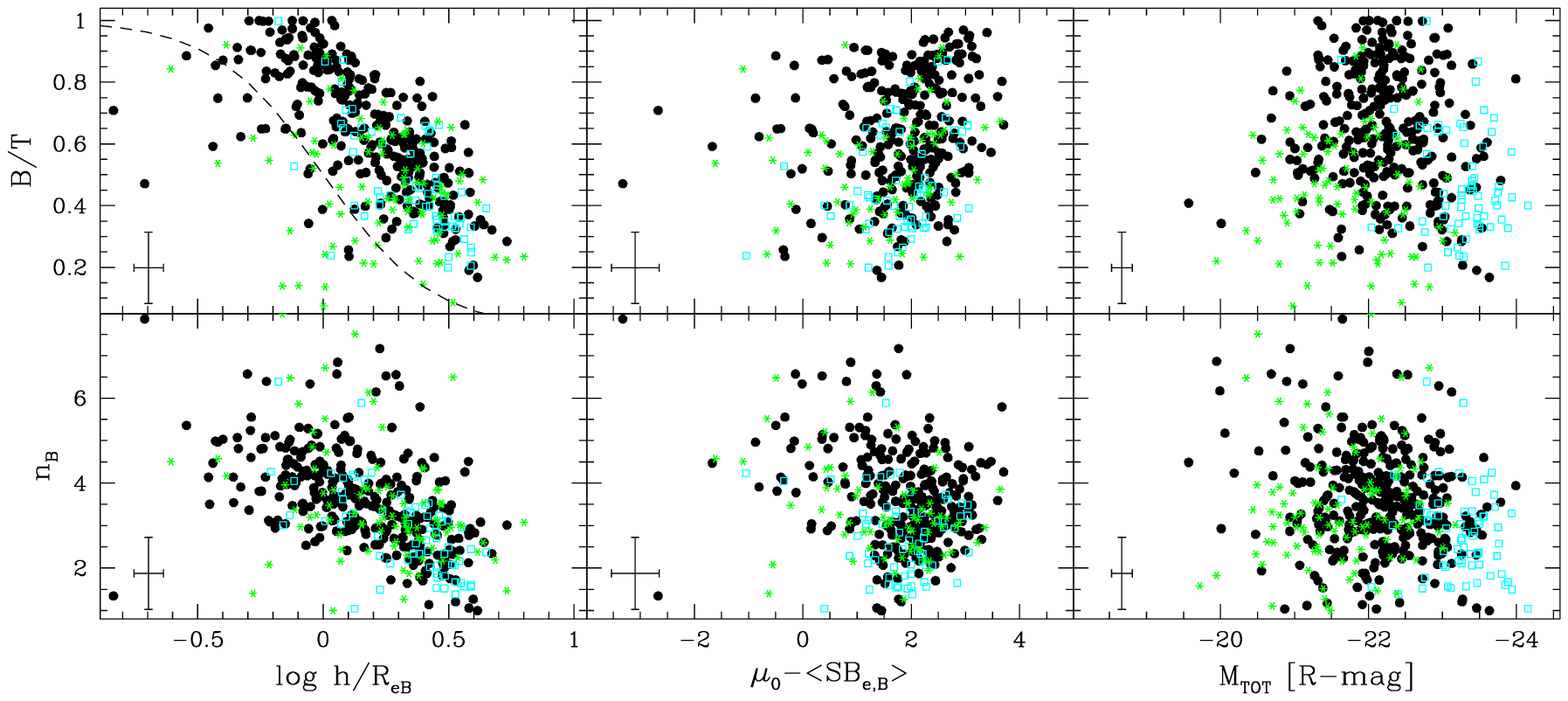}
\caption{Bulge-to-total-light ratio and S\'ersic profile shape parameter
  \nb\ as function of the bulge parameters (top), disc parameters
  (middle), and bulge-disc combined parameters (bottom).  Symbols and
  errorbars as in Fig.\,\ref{cmpMtotRe}. The dashed line in the
  $\h/\ReB$ versus \bt\ diagram indicates the expected relation
  between these parameters under the assumption that there is a fixed
  bulge to disc surface brightness difference with arbitrary zero-point.
}
\label{lgdbnb}
\end{figure*}

%
%

In Fig.\,\ref{lgdbnb} we show in three sets of diagrams the bulge,
disc, and combined bulge and disc parameters as function of \bt\ ratio
and S\'ersic \nb\ values. Before we can draw any conclusions on any
apparent correlation, we have to be aware of the many selection
effects playing in these diagrams. Some are explicit, like the limits
we placed on some parameters based on \bt\ ratio to reduce fit
errors. Others are more hidden, like the exclusion of bright discs, as
their more visible spiral structure would have precluded them from
entering the sample in the first place.

Concentrating first on the \bt\ ratios, we see that more bulge
dominated systems have larger bulge scale sizes, but lower effective
surface intensities, which combines in a rather weak \bt\ ratio trend
with total bulge luminosity. On the other hand, looking at the disc
parameters in the middle set of diagrams, we see a weak trend of disc
scale sizes with \bt\ ratio, but at each \bt\ ratio there is a wide
range in disc scale sizes. More bulge dominated systems have discs of
lower surface intensities. The combined trends of disc scale size and
surface brightness result in a strong correlation between disc
luminosity and \bt\ ratio for our sample dominated by bright, early-type
galaxies.

Combining these results for this sample, the change in \bt\ ratio
seems to be mainly driven by changes in disc luminosity, not bulge
luminosity. This is opposite to what has been observed for disc
dominated systems as found by e.g.\ \citet{Tru02} and \citet{Bal04},
who showed that most of the change in \bt\ ratio in these systems is
due to bulge luminosity variation. We have to be cautious though. Our
selection criteria resulted in a very limited range in total galaxy
luminosities. We are only sampling the upper decade in the luminosity
function, the 5--10 brightest galaxies in a cluster. It could be that
lower luminosity elliptical galaxies reveal often equal sized discs,
changing the observed trends.

Finally, we see that the weak opposite trends in the bulge and disc
scale sizes with \bt\ ratio combine in a strong correlation between
\h/\ReB\ and \bt. We should be cautious not to see too much in this
relationship. The expected relation between $\h/\ReB$ and \bt\ under
the assumption that there is a fixed bulge to
disc surface brightness difference and no significant bulge profile
shape change is indicated by the dashed line. The bulk of the galaxies
have indeed a limited range in disc to bulge surface brightness
difference as shown in Fig.\,\ref{lgdbnb}. It is just hard to see a
faint component on top of a bright component, unless their light
distributions are completely different. The variation from $\nb=1$ to
$\nb=5$ amounts to only a factor 2 in bulge luminosity change at fixed
bulge to disc surface brightness ratio (see \citealt{Gra01}, his fig.\,4)
and this change is systematic with \h/\ReB\ as shown in
Fig.\,\ref{lgdbnb}. All in all, the relation between \bt\ and \h/\ReB\
with the observed scatter is expected, given the observed scatter in
disc to bulge surface brightness contrast and the limited and
systematic effect in \nb.

We note that \citet{DOn01} finds a different slope in the \bt\ versus
\h/\ReB\ relation. Many of his discs were fitted to
the small inner components (either discs or cores) of the galaxies in
his nearby sample. We are not able to resolve these components and, as
D'Onofrio rightly indicates, it is unlikely that these inner
components represent similar structures as the more extended
components fitted in discy galaxies. We do fit the more extended
components and indeed find a they have different relation to the main
galaxy.


If we now turn our attention to the relationships between \nb\ and the
bulge and disc parameters, we note that \nb\ has a behavior very
similar to the \bt\ ratio with respect to these bulge and disc
parameters. There are some more outliers in the \nb\ relations, most
of which are for \nb\ values larger than 5.5. 

\begin{figure}
\epsfxsize=8.2cm
\epsfbox{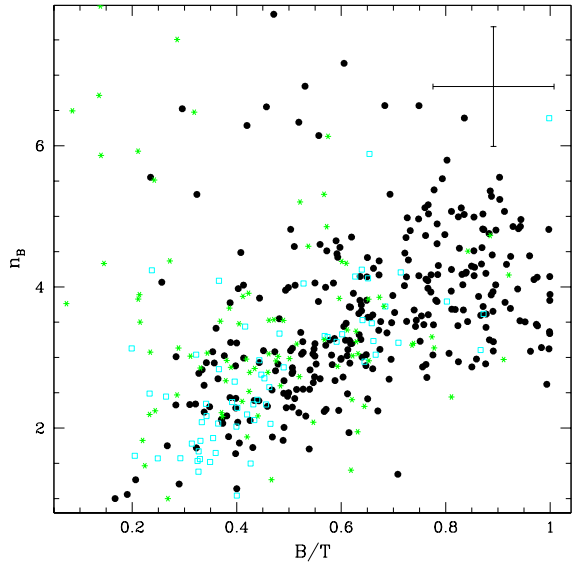}
\caption{The S\'ersic \nb\ bulge shape parameter versus \bt\ ratio. Symbols
and errorbars as in Fig.\,\ref{cmpMtotRe}.
}
\label{bt_nb}
\end{figure}

The very similar behavior of \nb\ and \bt\ ratio in Fig.\,\ref{lgdbnb}
suggests that these two parameters must be correlated, which is indeed
the case as shown in Fig.\,\ref{bt_nb}. As mentioned before, all other
things being equal a change in \nb\ from 1 to 5 only results only in a
factor 2 change in bulge to disc luminosity, so the change in
\nb\ does not drive the variation in \bt\ ratio on its own. For this
set of early-type galaxies, a large fraction of the \bt\ change seems
to come from the change in bulge effective radius, and it is the
correlation between \nb\ and \ReB\ that seems to drive the trend
between \nb\ and \bt\ ratio. Still, cause and effect are hard to
disentangle based on this photometric data set alone. We note that
these trends cannot solely be the result of the correlated errors
noted in Fig.\,\ref{testg2d}, as the slope of the errors is incorrect
and the size of the effect is too small.

\begin{figure}
\epsfxsize=8.2cm
\epsfbox{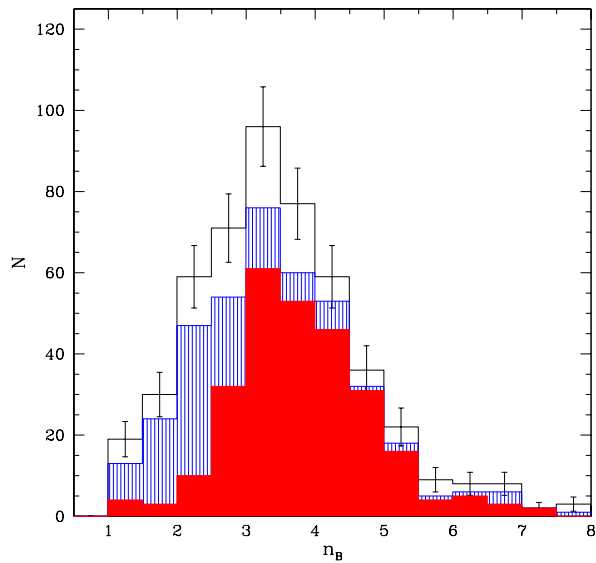}
\caption{
The distribution of S\'ersic profile shape parameter \nb\ for our
sample. The top solid line shows the total sample, the hashed
histogram shows all galaxies without clear spiral structures
and the solid histogram show all non-spiral galaxies with
$\bt>0.5$.
}
\label{nbhis}
\end{figure}

In Fig.\,\ref{nbhis} we show the observed distributions of the
\nb\ parameter for different subsets of our sample. Galaxies that were
fitted with \nb\ values larger than 6 were outliers in previous
diagrams and are considered of dubious quality. They are excluded from
further analysis here. The distributions are quite narrowly peaked,
with 1 sigma dispersions ranging from about 1.05 for the total sample
to 0.88 for bulge dominated non-spiral galaxies. This width is about
twice the mean of the individual maximal error estimates of the
\nb\ values, and is therefore significant. The median \nb\ value
shifts from 3.24 for the total sample, to 3.31 for the non-spiral
sample, and to 3.66 for the non-spiral, bulge dominated $\bt>0.5$
sample. For galaxies with $\bt>0.7$ the median is 3.99. 
The real \nb\ values might be even somewhat lower. \citet{Bal03}
showed that for a sample of early type disc galaxies the \nb\ values
decreased compared to ground based data when a nuclear component
revealed in their HST data was included as a seperate component in
their fits.

This result complicates bulge-disc decompositions of higher redshift
galaxies with typically lower signal-to-noise and resolution. On such
lower quality data a S\'ersic-law bulge with exponential disc does not
reliably converge and one is force to use a fixed \nb\ value. Given
that in this morphologically preselected and bright galaxy sample
there is a broad distribution of \nb\ independent of luminosity (see
Fig.\,\ref{lgdbnb}) peaking at values below \nb=4, and given that
spiral galaxies tend to have bulges with \nb\ values in the 0.5--3 range
(e.g., \citealt{Mac03, Bal03}), it may be more appropriate to use a \nb\ in
the 1.5--3 range rather than the more customary \nb=4 in fixed
\nb\ fits. A similar conclusion was reached for early-type disc galaxies
by \citet{Bal03}, but is now extended to bright, early-type
cluster galaxies here.

\subsection{Space density distribution functions}
\label{sec:space-dens-distr}

Probably more important than correlations between structural
parameters is the true space density of galaxies as function of
structural parameters, i.e.\ the distribution of structural parameters
corrected for selection effects.

When calculating space densities of galaxies from the EFAR sample
there are several points to note. The sample is not complete for
spiral galaxies, so we can not say anything about space densities of
those galaxies. The regions selected by EFAR where particularly rich in
clusters, and therefore are not representative for the average space
density of galaxies in the local universe. Assuming that galaxies have 
similar properties in different clusters, we can still use the
relative space density of structural parameters, but for absolute
space densities we will have to re-normalize the EFAR regions to the
universal cluster density.

Calculating the space density of galaxies requires good understanding
of the selection function. The selection of the EFAR sample is
described in detail in Paper I. The selection of EFAR galaxies was a
two step process, where first clusters were selected, believed to be
in the right redshift range and to be rich enough to obtain accurate
distances. In the next step galaxies were selected in each cluster
according to a minimum angular diameter criterion.

We will use here a simple \vmax\ correction method \citep{Sch68} to
calculate the relative space density of E/S0 galaxies as function of
their structural parameters. The \vmax\ correction method assumes that
the average space density of an object with certain parameters is
inversely proportional to the maximum ``visibility'' volume \vmax\ of
that object, given the selection criteria of the sample (for more
detailed descriptions see \citealt{Fel76, DisPhi83, deJLac00}). 
The selection of galaxies was a two step process and
therefore calculating the maximum visibility volume of a galaxy must
reproduce this two step process.

First, the \vmax\ of each individual galaxy was calculated, by
integrating the total volume occupied between the minimum and maximum
selection redshifts of $z_{\rm min}$=6000/$c$ and $z_{\rm
max}$=15000/$c$, with $c$ the speed of light. While doing the volume
integral, we have to take into account the
diameter selection function determined in Paper I for each cluster as applied
to each galaxy individually. The selection function describes the
selection diameter ($D_W$) dependent probability of sample inclusion,
parametrized by:
\begin{equation}
P(\log(D_W)) = 0.5(1+{\rm erf}[\log(D_W/D_W^0)/\delta_W]),
\label{D_W}
\end{equation}
where $D_W^0$ is the midpoint and $\delta_W$ the width of the cutoff
in the selection function. We furthermore took into account that the
contribution to the volume has to drop to zero when \Reff\ becomes
less than 1.5 arcsec had the galaxy been at that redshift, because we
excluded such galaxies from our analysis. The
integral to calculate the total visibility volume is then:
\begin{equation}
\vmax = 4\pi \Omega_f \int_{z_{\rm min}}^{z_{\rm max}} P(\log(D_W\frac{D_M(
z_{\rm Cl})}{D_M(z)}) ) \frac{c D^2_M(z) S(z)}{(1+z)H(z)}\,dz
\end{equation}
with
\begin{equation}
S(z) = \left\{ \begin{array}{ll}
	1 & \mbox{if $\Reff \frac{D_M(z_{\rm Cl})}{D_M(z)}\ge1.5$ arcsec}\\
	0 & \mbox{otherwise}
	\end{array}
	\right\},
\end{equation}
$z_{\rm Cl}$ the cluster redshift of the galaxy, $\Omega_f=0.139$, the
fraction of the sky covered by our survey, and $D_M(z)$ comoving distance
and $H(z)$ Hubble constant at redshift $z$ as defined in \citet{Hog99}.

In the second step, we have to take into account that only objects
were selected if they belonged to a cluster, with at least three
early-type galaxies obeying the diameter selection
criterion. Therefore, the \vmax\ of a cluster is identical to the
\vmax\ of the galaxy ranked third in diameter belonging to the
cluster. The final \vmax\ of each galaxy is the minimum of its
individual and cluster \vmax. This last step only modifies the \vmax\
of the top 3 diameter galaxies in each cluster, and only if the
diameter limit of the cluster is reached before the third ranked
galaxy is redshifted to 15000\,km/s.

As mentioned before, the EFAR regions were chosen to be particular
rich in clusters at the target redshift range. We used NED to estimate
the size of the over density. NED lists 140 original, non supplement
Abell clusters \citep{Abe58, Abe89} for $|b|>30\degr$
($\Omega_f=0.5$) with $6000<v<15000$\,km/s versus 87 in the EFAR
regions ($\Omega_f=0.139$). This translates in an over density of
about 1.75. However, we observed only 32 of the 87 clusters identified
by NED, and while not every NED cluster should be in our
sample (some cluster redshifts are based on only one galaxy redshift),
it is clear that the over density is probably offset by
incompleteness. Our sample contains another 41 clusters selected by
other means, which are not included in the Abell catalog. We have no
means to access the completeness of these in general smaller
clusters. Given these uncertainties, we have decided not to correct
for over density and incompleteness, and assume that the uncertainty
in the absolute space density zero-point is a factor of two. We do
however correct for our ellipticity incompleteness, assuming that
elliptical galaxies with \ellip$<$0.4 constitute 65\% of the total
E+S0 population \citep{JorFra94}.

We will show the factor 2 uncertainty in absolute zero-point of the space
density distributions in the diagrams presented in the remainder of
this section. This uncertainty will not affect the conclusions drawn
from these diagrams. We will also show that our space density
distribution and normalization is remarkably consistent with that of
the SDSS, giving confidence that our normalization is indeed correct
within a factor of 2.

\begin{figure}
\epsfxsize=\linewidth
\epsfbox[20 145 482 700]{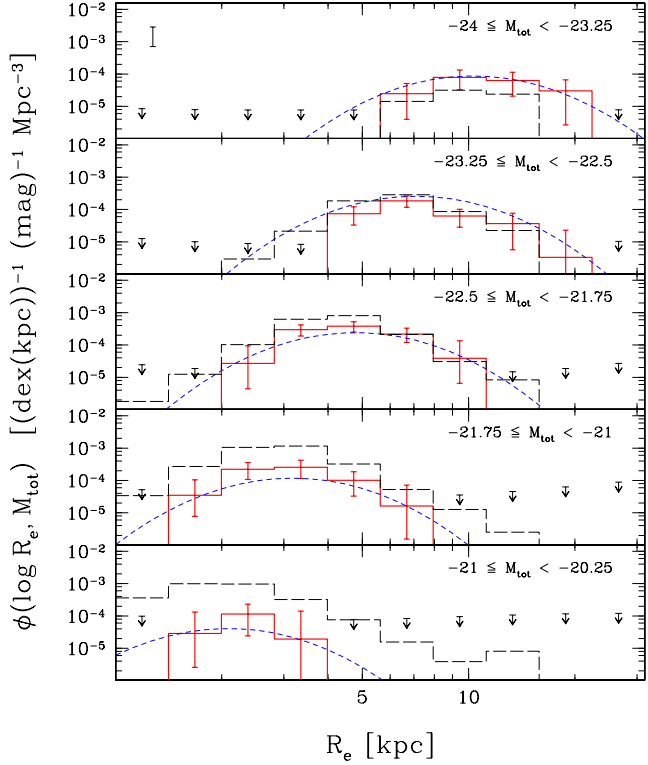}
\caption{The space density of all non-spiral galaxies as function of
total magnitude and effective radius of the whole galaxy is
shown as the solid line histogram. Errorbars on the histogram show the
95\% confidence limits due to uncertainties in selection diameter and
distance and due to Poisson statistics. The upper limit symbols
indicate the 95\% confidence limits based on the non-detection of such
galaxies and the visibility volume of \Rq\ profiles shape galaxies for
our selection criteria. The errorbar in the top-left corner of the top
panel shows the factor 2 uncertainty in the absolute space density
normalization of our distributions. The dashed curve shows the bivariate
distribution function of the form presented in \citet{deJLac00}
fitted to the data. The dashed histogram is the space density
distribution derived by \citet{Bla03} from SDSS data
for galaxies with $n>3$ and scaled as described in the text.}
\label{dismag_lgre}
\end{figure}

In Fig.\,\ref{dismag_lgre} we show the combined bivariate space
density of all non-spiral galaxies as function of absolute magnitude and
effective radius. The 95\% confidence limit error estimates on the
number density distribution and the 95\% confidence upper limits in
the regions with no galaxies in the sample were derived in a similar
fashion as described in \citet{deJLac00}, assuming a perfect \Rq\
law luminosity profile for all galaxies. These error estimates and
upper limits include the effects of distance and diameter
uncertainties and Poisson statistics.

Figure\,\ref{dismag_lgre} also shows the fitted bivariate distribution
function of the form described by \citet{deJLac00}. This function
takes the shape of a \citet{Sch76} luminosity function in the
luminosity dimension (with the usual parameters $\phi_*$, $M_*$, and
$\alpha$), and of a log-normal distribution in the scale size
dimension, with a width $\sigma_\lambda$, a median scale size for
galaxy with $M=M_*$ of $R_{e*}$, and a shift in median scale size with
luminosity parametrized by $\beta$:
\begin{eqnarray}
 \label{bivareq}
 \lefteqn{\phi(M,\log(\Reff))\,dM\,d\log\Reff = 0.4\ln(10)\ \frac{\ln(10)}{\sqrt{2\pi} \sigma_\lambda}}  \\
 & \times \phi_*\ 10^{-0.4(\alpha+1)(M-M_*)} \,
     \exp(-10^{-0.4(M-M_*)}) \,dM \nonumber \\[1.3mm]
 & \times   
     \exp\left[ -\frac{1}{2}\left( \frac{\log(\Reff/R_{e*})-0.4\beta(M-M_*)}
     {\sigma_\lambda/\ln(10)} \right)^2 \right] \, d\log\Reff. \nonumber
\end{eqnarray}
The second line in this equation represents the Schechter luminosity
function in magnitudes, while the third line represents the log-normal
scale size distribution.

The function was fitted using a maximum likelihood technique and the
resulting parameters are listed in Table\,\ref{bidistab}, with the
errors indicating the 95\% confidence limits. These confidence limits
were determined by means of Monte Carlo bootstrap resampling of the
galaxies \citep{Pre93}, where we also varied the size of the bins in
both directions. The fitted function is a good description of the
observed distribution: the goodness-of-fit parameter $Q$ \citep{Pre93}
was larger than 0.1 in more than 28\% of the bootstrap resampled
realizations and $Q>0.001$ in more than 87\% of the realizations.  A
function similar in shape was fitted by \citet{Cho85} to a sample of
233 elliptical galaxies and our parameters do agree to within the
quoted uncertainties.

\begin{table*}
 \caption{Bivariate distribution function parameters. Errors indicate 95\% confidence limits.} 
 \label{bidistab}
\begin{tabular}{@{}lcccccc}
\hline
\hline
\multicolumn{1}{c}{Fit}   & $\phi_*$ & $\alpha$ & $M_*$ & $\Reff$$_*$ & $\sigma_\lambda
$ & $\beta$\\
       &($\times 10^{-4}$\,Mpc$^{-3}$)& & ($R$-mag) & (kpc) & & \\
\hline
Total Galaxy& 0.99 $\pm$ 0.19 &  1.06 $\pm$ 0.26 & -21.81 $\pm$ 0.18 & 3.9 $\pm$ 0.5 & 0.34 $\pm$ 0.04 & -0.62 $\pm$ 0.08\\
Bulge       & 0.83 $\pm$ 0.31 &  1.30 $\pm$ 0.50 & -21.15 $\pm$ 0.30 & 2.3 $\pm$ 0.4 & 0.39 $\pm$ 0.04 & -0.55 $\pm$ 0.10\\
Disc        & 0.91 $\pm$ 0.26 & -0.17 $\pm$ 0.24 & -21.98 $\pm$ 0.35 &12.6 $\pm$ 2.2 & 0.42 $\pm$ 0.04 & -0.40 $\pm$ 0.07\\
\hline
\end{tabular}
\end{table*}

In Fig.\,\ref{dismag_lgre} we compare our bivariate distribution with
the findings of \citet{Bla03}, based on the SDSS data (see their
Fig.\,13). They fitted S\'ersic profiles to the total galaxy
luminosity profiles, without an added exponential disc component and
hence without doing a full bulge/disc decomposition. We show their
$^{0.1}i$ distribution for all galaxies with an S\'ersic $n>3$. We
scaled their distribution to our Hubble constant and using a $^{0.1}i$
to $R$ conversion of 0.2 mag (from \citealt{Bla03}
$^{0.1}(r-i)$$\sim$0.4, $^{0.1}r$ = $^{0.0}r-2.5\log(1.1)$, and $R\sim
r-0.3$ from \citealt{BelVan87}).  

Surprisingly, the distributions agree well at the bright end, even
though our absolute space density zero-point calibration is somewhat
uncertain (see errorbar in top panel) due to the cluster over density
in the EFAR region and our limited knownledge of cluster
incompleteness. The cutoff at \Reff$>$20\,kpc in the \citet{Bla03}
distribution is not real, but due to the cutoff in their figure. At
fainter luminosities our distribution is lower than that of
\citet{Bla03}. One explanation for this might be that we were stricter
in excluding spiral galaxies from our distribution, while the Blanton
et al.\ distribution includes all galaxies with $n>3$, which may
include spiral galaxies whose combined bulge and disc luminosity
profile mimics that of an elliptical galaxy. On the other hand,
incompleteness may affect our distribution at the faint end. Our
sample was drawn from rich clusters, while the \citet{Bla03} SDSS
distributions were calculated for all sky. This will have little
effect on the bright end of the distribution, because the brightest
elliptical galaxies are only found in cluster, but will affect the
faint end. Notwithstanding, the size distributions at a given
luminosity are very similar.

\begin{figure}
\epsfxsize=\linewidth
\epsfbox[20 145 482 700]{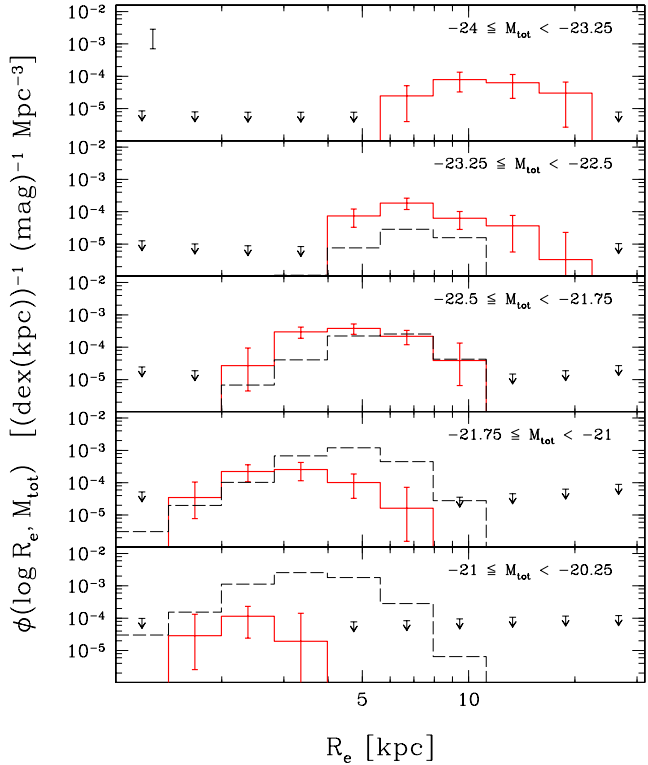}
\caption{The space density of EFAR galaxies as function of magnitude
  and effective radius of the total galaxy (solid
  histogram). Errorbars and upper limits the same as in
  Fig.\,\ref{dismag_lgre}. The dashed histogram is the space density
  distribution derived by \citet{Bla03} from SDSS data for galaxies
  with $n<1.5$ and scaled as described in the text. }
\label{dismag_lgre_spir}
\end{figure}

In recent years it has become clear there are two types of elliptical
galaxies \citep{KorBen96,Fab97}. On the one hand we have the bright
elliptical galaxies with boxy isophotes, with a clear central core,
dynamically mainly pressure supported, while on the other hand we have
the lower luminosity elliptical galaxies, often with more discy isophotes,
cuspy cores and for a larger fraction supported and flattened by
rotation. The boxy galaxies tend to be central dominant galaxies of
clusters (cD galaxies) or at the center of one of the subclumps of a
cluster (resulting from cluster merging?). The transition from one
type into the other occurs at about -21.2 $V$-mag \citep{Fab97},
i.e.\ about -20.5 $R$-mag, with an overlap of both types occurring for
about 1.5 mag.

The bright, boxy type elliptical galaxies are probably the result of
the merging of many subclumb units in the hierarchical galaxy
formation scenario. The origin of the lower luminosity elliptical
galaxies is less clear. They could be the result of the merging of two
similar sized gaseous galaxies \citep{Bar92} or the result of
``failed'' spiral galaxies due to the lack of angular momentum in the
proto-galaxy (e.g., \citealt{Dal97,deJLac00}). These two options might
actually be somewhat related, as merging of two sub units is of course
the result of lack of orbital angular momentum. For S0 galaxies even
more formation scenarios have been proposed: ram-pressure stripping of
gas \citep{GunGot72,Ken81,Qui00}, galaxy harassment via high speed
impulsive encounters \citep{Moo96}, cluster tidal forces
\citep{ByrVal90} which distort galaxies as they come close to the
centre, interaction/merging of galaxies \citep{Ick85, Bek98}, and
removal \& consumption of the gas due to the cluster environment
\citep{Lar80, Bek02}.  In all these
scenarios the origin of non-boxy elliptical galaxies and S0 galaxies
lies in spiral galaxies or proto-spiral clumps. Therefore, it seems
natural to compare the size-luminosity distribution of our ellipical
sample to that of spiral galaxies.

In Fig.\,\ref{dismag_lgre_spir} we show again the space density
distribution of EFAR galaxies as function of size and luminosity, but
this time we compare the distribution with that of \citet{Bla03}
SDSS galaxies with S\'ersic profile fits with $n<1.5$, i.e.\ galaxies
with exponential-like profiles, predominanlty spiral galaxies. The
Blanton et al.\ distributions were scaled for Hubble constant and
passband in the same way as in the previous diagram for $n>3$ galaxies.
The disc effective radii were multiplied by 0.75 to take into
account that for an \Rq-law galaxy the projected surface brightness
effective radius is about 0.75 the spherical effective radius. In this way,
we make a rough comparison between the stellar mass distributions,
ignoring the effect of other baryonic components (most notably the
\hi\ in the disc galaxies, which could be converted to stars during
merging) and the systematic change in $M/L$ with luminosity,
especially for disc galaxies. When scaled for Hubble constant and
passband, the Sc-Sd spiral galaxy distribution of \citet{deJLac00}
is very similar to the one of \citet{Bla03}, particularly at the
bright end.

Figure\,\ref{dismag_lgre_spir} immediately shows that to form the
brightest elliptical galaxies one will need to merge several current
day bright spiral galaxies, because there is a substantial space
density of bright elliptical galaxies that are at least 0.75 magnitude
brighter than the brightest spiral galaxies. In the process the
effective radius of the resulting galaxy is increased by a factor of
few compared to the originating galaxy, presumably because the
kinematic energy of the galaxies orbiting each other is converted to
heat the random kinematic energy of the stars in the merger
product. \citet{Bar92} finds for dissipationless equal mass mergers a
typical increase of effective radius of about 1.4, similar to what one
expects assuming conservation of energy and the virial theorum 
\citep{Col00,She03}. We have stepped our luminosity bins in
Fig.\,\ref{dismag_lgre_spir} by 0.75 mag, i.e. about a factor 2 in
luminosity (and roughly stellar mass), so each equal mass
dissipationless merger of two disc galaxies would result in a merger
remnant one bin up with about 1.4 larger size. Merging of at least a
few and probably many more of present day spiral galaxies would be
needed to form the brightest elliptical galaxies. Because merging of
spiral galaxies will involve dissipation and therefore the scale size
may be less, it may be more reasonable to think that the brightest
elliptical galaxies formed from the merging of several smaller E/S0
galaxies.  

A similar conclusion was reached by \citet{She03}. They showed that by
having a series of mergers of small elliptical galaxies they could
create the observed size-luminosity relation and its dispersion, when
making reasonable assumptions about the transfer of orbital angular
energy. However, they did not compare space densities, and furthermore,
this process leaves the question of the origin of small elliptical
galaxies unanswered.

The distribution of the lower luminosity elliptical galaxies
(\mtot$>$-22.5 $R$-mag) in Fig.\,\ref{dismag_lgre_spir} is more
surprising. Even though in number density not completely unlike the
spiral galaxies (remember the absolute zero-point for the elliptical
distribution is somewhat uncertain), the effective radius distribution
of the elliptical galaxies peaks at smaller radii than the
distribution of disc galaxies. As argued before, dissipationless
merging results in larger scale sizes, typically by a factor of about
1.4, and therefore the dissipationless merging of typical current day
spiral galaxies cannot result in these current day lower luminosity
elliptical galaxies.

What other creation scenarios can we invoke to explain the lower
luminosity elliptical galaxies? First of all, the merging of two
spiral galaxies is not going to be dissipationless. Simulations that
include gas show that during the merger process gas will quickly
stream to the center of the merging galaxies, presumably creating a
starburst and dragging along some of the existing stellar population,
creating a more concentrated stellar remnant. Whether this process is
enough to offset the effect of the added orbital kinematic energy is
questionable, given that current bright spiral galaxies have gas mass
fractions of 5--20\%. At higher redshifts the gas fractions will have
been higher and therefore this process may have been more efficient
then. In the hierarchical galaxy formation picture, galaxies at higher
redshifts have smaller scale sizes, so the small scale sizes of low
luminosity elliptical galaxies in clusters may be the result of high
redshift mergers which have subsequetially not been able to grow any
more in the cluster environment.

Another option to explain the small scale sizes of lower luminosity
elliptical galaxies might come from the failed disc scenario proposed
by several authors for different reasons (e.g.,
\citealt{Dal97,Mo98,deJLac00,McGdeB98}). \citet{deJLac00} found that the scale
size distribution of spiral galaxies used here was narrower than what
one would expect in hierarchical galaxy formation scenario where disc
galaxy sizes are determined by their spin acquired from tidal torques
with neighbors. They proposed (similar to \citealt{Mo98}) that the lowest
angular momentum proto galaxies never were able to form a proper disc
and immediately collapsed to spheroids. The problem here is that we
cannot compare the elliptical galaxy distribution drawn from a cluster
sample to the distribution of spiral galaxies drawn mainly from the
field.

Can spiral galaxies falling into a cluster result in a elliptical
scale size distribution as observed here? Simple stripping of gas by
ram-pressure or tidal interactions and subsequent fading of the stellar
population is clearly not enough, elliptical galaxies are
just too small compared to spiral galaxies. Subsequent stripping of
the stars in the outer parts of the infalling disc galaxy by galaxy
harassment \citep{Moo96} could result in smaller effective radii of the
remnants. Still a lot of stripping would have to occur, as stripping
half the stellar mass exclusively from the outer part of the galaxy
would shift the disc distribution one bin down while dividing the size 
by two. A considerable fraction of this process would have to occur at 
quite high redshifts ($z>1.5$) as only then bright disc galaxies were
substantially smaller than current day disc galaxies \citep{deJLac99,Rav04,Fer04}.

As so often, several of the options listed above combined may be the
final answer. However, these relative elliptical and spiral galaxy
scale size distributions suggest that a substantial fraction of the
lower luminosity elliptical galaxies were created at quite high
redshift, not only in stars but also in structural parameters, when
disc galaxies were still smaller and more gas rich.

\begin{figure}
\epsfxsize=\linewidth
\epsfbox[20 145 482 700]{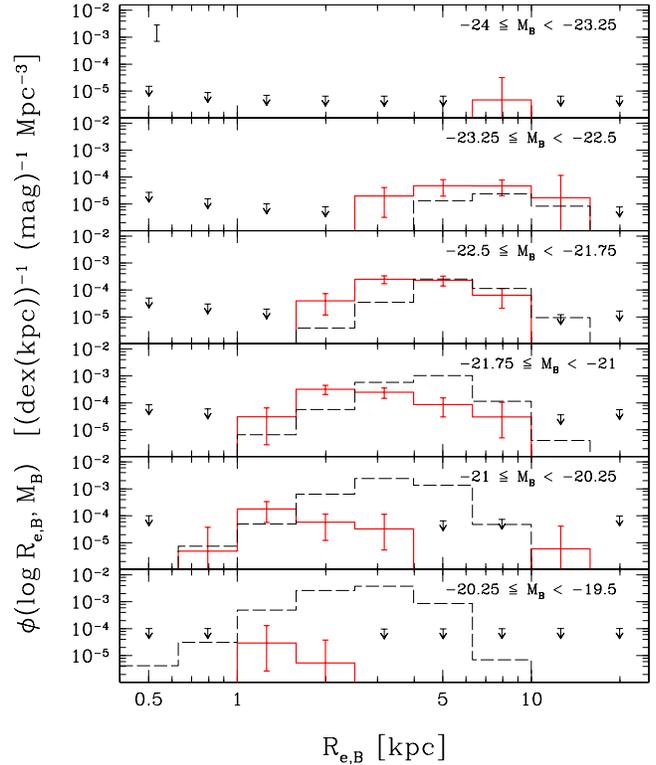}
\caption{The space density of EFAR elliptical galaxies as function of 
bulge magnitude and bulge scale length (solid histogram). Errorbars
and upper limits are the same as in Fig.\,\ref{dismag_lgre}. The
dashed histogram is the space density distribution derived by \citet{Bla03}
from SDSS data for galaxies with $n<1.5$,
scaled as described in the text.}
\label{dismagb_lgreb_spir}
\end{figure}

\begin{figure}
\epsfxsize=\linewidth
\epsfbox[20 145 482 700]{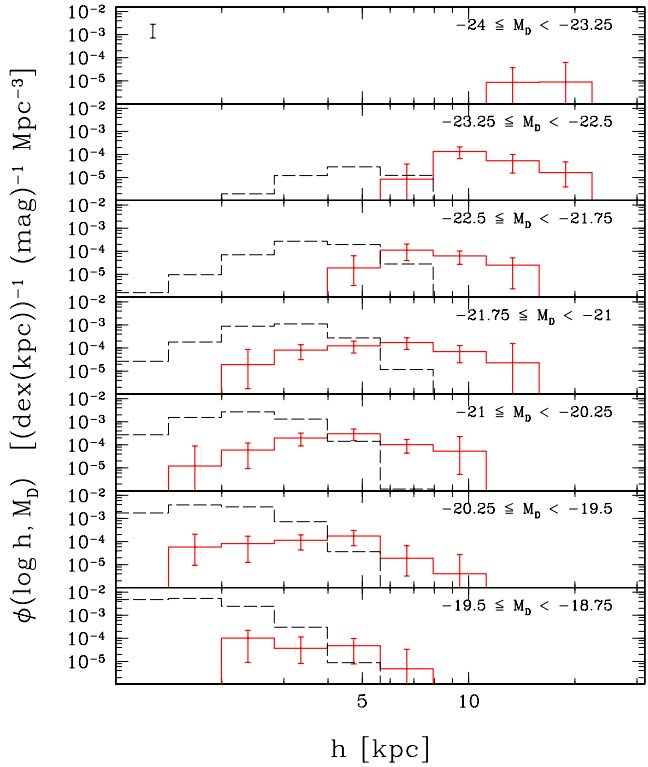}
\caption{The space density of EFAR
galaxies as function of disc magnitude and disc scale
length. Errorbars the same as in Fig.\,\ref{dismag_lgre}.  The dashed
histogram is the space density distribution derived by \citet{Bla03}
from SDSS data for galaxies with $n<1.5$,
scaled as described in the text.}
\label{dismagd_lgh_spir}
\end{figure}

Until so far, we have only looked at the luminosities and sizes for
the total systems, but in Figs.\ref{dismagb_lgreb_spir} and
\ref{dismagd_lgh_spir} we show the luminosity-size distributions for
the bulges and the discs, respectively. The luminosity and size parameters
are much better determined for the total system than for its components, but
nontheless it will be instructive to compare the distributions of the
components to that of spiral galaxies.

Comparison of the elliptical bulge parameters with the SDSS disc
parameters in Fig.\,\ref{dismagb_lgreb_spir} shows that the
distributions are somewhat similar, but that the bulges are about a
factor $\sim$1.5 to 2.5 smaller than the spiral discs from the
brightest to the faintest luminosities. The scaling of the SDSS disc
parameters is here the same as in Fig.\,\ref{dismag_lgre_spir},
including the factor 0.75 decrease in disc scale sizes to account for
the difference of a spherical bulge projected to a 2D effective radius
versus a 2D disc effective radius.

On the other hand, when we compare in Fig.\,\ref{dismagd_lgh_spir} the
SDSS $n<1.5$ disc parameter distribution with the distribution for the
component identified as discs in our EFAR sample, we see again
somewhat similar, parallel distributions, but this time the EFAR
disc-like component is about a factor two larger than the SDSS discs.
We have to be careful, because as argued before in
section\,\ref{sec:struct-param-corr} when discussing
Fig.\,\ref{lgh_mu0}, discs with low surface brightness and/or small
scale length might be hard to detect when superposed on a high surface
brightness bulge and therefore this distribution may be incomplete.
Still, it is clear that there are larger disc components in the EFAR
non-spiral galaxies than seen in the SDDS spiral discs. Upper limits
are somewhat hard to draw on this diagram, as the selection was
dominated by the bulge light distribution, not by the disc light.

Without kinematic information proving that the B/D decompositions we
made of our sample has an underlying physical meaning it is somewhat
premature to interpret these distributions in detail, but we can make
some speculations. Could the spheroidal components be the result of
early, disipationless mergers of roughly equally sized galaxies from
the time when galaxies were still much smaller ($z$$>$1--2)? The
disc-like component can clearly not be formed from normal disc
galaxies, there scale sizes are just too large for their
luminosity. The disc-like component tends to dominate a bit in the
most luminous galaxies (see Fig.\,\ref{lgdbnb}). Are these the stars
from smaller, infalling galaxies which maintain a larger component
from their initial orbital energy? Do they form disc-like components
because they predominantly fall into the cluster on the central galaxies
along the filamentary structures of the cosmic web?

While attractive from the galaxy structure point of view, this picture
leaves some other properties of elliptical galaxies
unexplained. Elliptical galaxies in clusters obey a colour-magnitude
relation up to redshifts of $\sim$0.9 (e.g., \citealt{Sta98}), which
means that this relation is driven by a mass-metallicity relation,
bigger galaxies being more metal rich. This therefore means that the
proto-galaxies that formed the big spheroids in the above picture
would have to have higher metallicity before they merged than the
proto-galaxies of smaller galaxies. Big elliptical galaxies also have
higher $\alpha$-element over Fe abundance ratios than smaller
galaxies, meaning that their star formation time-scales must have been
shorter (e.g., \citealt{Meh03}). In the hierarchical galaxy formation
picture this could be accomplished by having the proto-galaxies that
will merge into big galaxies form early on top of an over-density of
what will later become a cluster of galaxies at redshifts much larger
than 3. In a $\Lambda$CDM Universe there would be a lot of early
merging causing vigorous and short time-scale star formation, leading
to non-Solar rates of $\alpha$-element over abundances. Because these
proto-galaxies would already live in the deeper potential well of what
will later become a large galaxy in a cluster of galaxies, they would
be better in holding on to their enriched gas and therefore be able to
create the mass-metallicity relation. However, in the above picture
galaxies falling in along the filaments would have lower
(over-)abundances and would potentially be somewhat younger, and hence
the disc components should have lower metallicity and age. While
\citet{deJDav97} show that discy elliptical galaxies indeed have line
indices indicating younger ages and/or lower metallicities, spatial
resolved line indices mapping as performed by for instance the SAURON
team \citep{Fal04} will be needed to fully access this model.

\begin{figure}
\epsfxsize=\linewidth
\epsfbox[20 145 482 700]{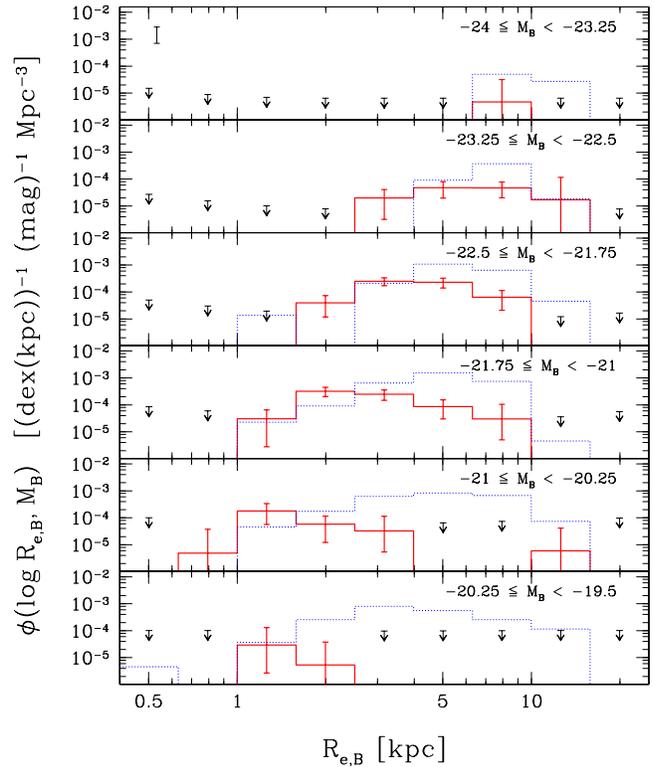}
\caption{The space density as function of bulge magnitude and bulge
  effective radius for the EFAR galaxies (solid histogram) and for the
  semi-analytic models of \citet{Col00} (dotted histogram). Errorbars
  and upper limits the same as in Fig.\,\ref{dismag_lgre}.}
\label{dismagb_lgreb_mod}
\end{figure}

Figure\,\ref{dismagb_lgreb_mod} shows the comparison of the (bulge
magnitude, bulge effective radius)-bivariate space density of the EFAR
elliptical galaxies with the models of \citet{Col00}. The
models are over-predicting the effective radii of early-type bulges by
about a factor of two, even more at fainter magnitudes. This is the
same problem as identified in Fig.\,\ref{dismag_lgre_spir}; the
effective radii of low luminosity early-type galaxies are smaller than
those of spiral galaxies of similar luminosity. Cole et al.\ assumed
that all early-type galaxies were created by merging in their models,
and as their models do a reasonable job in modelling the spiral galaxy
scale size distribution \citep{deJLac00}, the models will
over-predict the early-type galaxy scale size in the same way as was
discussed for Fig.\,\ref{dismag_lgre_spir}.

\section{Summary and conclusions}

We have performed 2D bulge/disc decompositions on a sample of 558
early-type galaxies from the EFAR sample using the GIM2D package. In
contrast with most earlier work, we have used a S\'ersic luminosity
profile for the spheroidal component, while using the common
exponential light profiles for the disc-like component. We showed
with extensive testing on model galaxies that the total galaxy
parameters and disc parameters can be recoverd at the 10-15\% level,
but that the bulge parameters are only recovered at the 30\% level,
mainly due to degeneracies resulting in correlated errors between \nb\
and \Reff. Our galaxy selection criteria are well known and we
have calculated bivariate space densities of elliptical galaxies as
function of galaxy parameters.

Without kinematic or population information it is impossible to prove
that our bulge and disc parameter are physically meaningful, but under
the assumption that the spheroid and disc parameters represent real
galaxy components we derive the following conclusions:

\begin{enumerate}

\item The scale sizes of the bulge and disc components are correlated.
  This may be partly due to the exclusion of spiral galaxies from the
  sample, excluding objects with small bulges and large discs.
  However, it should be noted that spiral galaxies show a similar
  trend \citep{deJ96III,Cou96,Mac03}.

\item The bulge-to-disc \bt\ ratios, the S\'ersic \nb\ bulge shape
  parameters, and the bulge effective radii show positive correlations
  with each other. We have no model that tells us which is cause and
  what is effect.

\item The median \nb\ value is 3.24 for all galaxies in our sample, is
  3.66 for all non-spiral, bulge dominated galaxies in our sample, and
  is 4 for non-spiral galaxies with \bt$>$0.7. Given that most spiral
  galaxies have bulge \nb\ values much lower than these values, this
  means that the standard \Rq\ de Vaucouleurs (i.e.\ \nb=4) fitted to
  spheroids may not be the most appropriate choice when doing
  bulge/disc decompositions on a random set of galaxies. In cases
  where one does not have the signal-to-noise or the spatial
  resolution to perform full S\'ersic bulge plus exponential disc fit,
  it will be more correct to use fixed value of \nb$\sim$3.5 for the
  brightest galaxies. Lower values should be used for less luminous
  galaxies.

\item The (luminosity, scale size) bivariate space density
  distribution of bright elliptical galaxies is well described by the
  analytic parametrization presented by \citet{deJLac00}.

\item Comparing the total galaxy (luminosity, scale size) bivariate
  distributions of EFAR non-spiral galaxies with SDSS disc-like
  galaxies shows that the brightest early-type galaxies could in
  principle be formed by merging a few large, current day spiral
  galaxies.  Low luminosity early-type galaxies have much smaller
  scale sizes than spiral galaxies of the same luminosity, and hence
  they can surely not be created from simple merging of two current
  day spiral galaxies.

\item When comparing the EFAR bulge and disc (luminosity, scale size)
  bivariate distributions to those of the SDSS disc-like galaxies, we
  showed that the bulges of early-type galaxies are typically a factor
  1.5 to 2.5 smaller and the discs a factor 2 larger than current day
  disc galaxies at a given luminosity.

\end{enumerate}

We speculate that the spheroidal components are the result of merging
of similar sized small proto-galaxies galaxies at high redshifts,
while the disc-like component may be the result of smaller galaxies
falling in later along the filaments of the cosmic web. While
attractive from the galaxy structure point of view, this model
requires careful tuning of the formation process of the proto-galaxies
at high redshift in order to reproduce the mass-metallicity relation
and the $\alpha$-element over-abundace observed at lower redshifts.


\section*{Acknowledgements}

We thank Frank Summers for generously making his computing resources
available for this project.  We thank an anomynous referee whos
comments have helped us to improve the paper. Support for RSdJ was
partly provided by NASA through Hubble Fellowship grant
\#HF-01106.01-A from the Space Telescope Science Institute, which is
operated by the Association of Universities for Research in Astronomy,
Inc., under NASA contract NAS5-26555. This work was partially
supported by NSF Grant AST90-16930 to DB, AST90-17048 and AST93-47714
to GW, AST90-20864 to RKM. The entire collaboration benefitted from
NATO Collaborative Research Grant 900159 and from the hospitality and
monetary support of Dartmouth College, Oxford University, the
University of Durham and Arizona State University. Support was also
received from PPARC visitors grants to Oxford and Durham Universities
and a PPARC rolling grant: ``Extragalactic Astronomy and Cosmology in
Durham 1994-98''.  This project made use of STARLINK facilities in
Durham.

This research has made use of NASA's Astrophysics Data System. This
research has made use of the NASA/IPAC Extragalactic Database (NED)
which is operated by the Jet Propulsion Laboratory, California
Institute of Technology, under contract with the National Aeronautics
and Space Administration.  

Funding for the creation and distribution of the SDSS Archive has been
provided by the Alfred P. Sloan Foundation, the Participating
Institutions, the National Aeronautics and Space Administration, the
National Science Foundation, the U.S. Department of Energy, the
Japanese Monbukagakusho, and the Max Planck Society. The SDSS Web site
is http://www.sdss.org/.

The SDSS is managed by the Astrophysical Research Consortium (ARC) for
the Participating Institutions. The Participating Institutions are The
University o f Chicago, Fermilab, the Institute for Advanced Study,
the Japan Participation Group, The Johns Hopkins University, Los
Alamos National Laboratory, the Max-Planck-Institute for Astronomy
(MPIA), the Max-Planck-Institute for Astrophysics (MPA), New Mexico
State University, Princeton University, the United States Naval
Observatory, and the University of Washington.


\end{document}